\documentclass[aip,jcp,floatfix,reprint]{revtex4-1}
\usepackage{amsmath}    
\usepackage{graphicx}   
\usepackage{verbatim}   
\usepackage{color}      
\usepackage{subfigure}  
\usepackage{soul}
\usepackage{epstopdf}

\newcommand{\teo}[1]{{\color{black}  #1}}

\newcommand{\teoqq}[1]{{\color{black}  #1}}
\newcommand{\teoqqq}[1]{{\color{black}  #1}}
\newcommand{\teoedit}[1]{{\color{black} #1}}
\newcommand{\teoeditnew}[1]{{\color {black} #1}}
\newcommand{\teoeditnewb}[1]{{\color {black} #1}}
\newcommand{\teoresp}[1]{{\color {black} #1}}

\begin{document}

\title{The thermodynamics of computational copying in biochemical systems}
\author{Thomas E. Ouldridge}
\affiliation{Department of Bioengineering, Imperial College London, London, SW7 2AZ, UK}
\email{t.ouldridge@impeial.ac.uk}
\author{Christopher C. Govern}
\affiliation{FOM Institute AMOLF, Science Park 104, 1098 XE Amsterdam,
  The Netherlands} 
  \author{Pieter Rein ten Wolde} \affiliation{FOM
  Institute AMOLF, Science Park 104, 1098 XE Amsterdam, The
  Netherlands}

\begin{abstract}
\teoeditnew{Living cells use readout molecules to record the state of  receptor proteins, similar to measurements  or copies in typical computational devices. But is this analogy rigorous? Can cells be optimally efficient, and if not, why? We show that, as in computation, a canonical biochemical readout network generates correlations; extracting no work from these correlations sets  a lower bound on dissipation. For general input, the biochemical network cannot reach this bound, even with arbitrarily slow reactions or weak thermodynamic driving. It faces an accuracy-dissipation trade-off that is qualitatively distinct from and worse than implied by the bound, and  more complex steady-state copy processes cannot perform better. Nonetheless, the cost remains close to the thermodynamic bound unless accuracy is extremely high. Additionally, we show that biomolecular reactions could be used in thermodynamically optimal devices under exogenous manipulation of chemical fuels, suggesting an experimental system for testing computational thermodynamics. }
\end{abstract}


\maketitle

If it were possible to perform many measurements using a single bit of memory without putting in work, Maxwell's Demon could use the information gained to violate the second law of thermodynamics and extract net work from an equilibrium system.
\teoeditnewb{Landauer's insight that computational processes require a physical instantiation and therefore have thermodynamic consequences \cite{Landauer1961, Bennett1982, Bennett2003} is key to exorcising the Demon, and the survival of the second law has been demonstrated in a range of physical models \cite{Bennett1982,Bennett2003,Sagawa2009}. If, unlike Maxwell's thought experiment, the correlations generated by a measurement or copy are not used to perform work,}  the cycle increases the entropy of the universe (by at least $k \ln 2$ if the measurement is binary, perfectly accurate and has a 50/50 outcome) \cite{Bennett1982,Sagawa2009,Fahn1996}. Landauer and others have provided specific physical implementations of binary devices along with protocols that achieve the thermodynamic bound for measurement cycles \cite{Bennett1982, Bennett2003} or memory erasure protocols \cite{Landauer1961, lambson2011exploring, berut2012experimental, Jun2014,Hong2014}. Examples include magnetic systems \cite{Landauer1961,Bennett1982,Hong2014}, or single particles in pistons \cite{Szilard1964, Fahn1996, Plenio2001, Ladyman2007, Maroney2005, Sagawa2009}.

Do biomolecules  perform measurement and copying
within this computational paradigm? Many biological processes involve creating long-lived
molecular copies of other molecules
\cite{Bennett1982,feynman1998feynman,Parrondo2015}. Perhaps the most
tantalising analogy is in the cellular sensing of external
ligand concentrations. Following the seminal work of Berg and
  Purcell \cite{berg1977}, it has been shown that cells
can reduce their sensing error by averaging a noisy receptor signal
over time \cite{bialek2005,levinepre2007,
  levineprl2008,wingreen2009,levineprl2010,mora2010,Govern2012,
  govern2014}. \teoedit{Recent studies claim that cells
 implement time integration by dissipatively copying 
receptor states into the chemical modification states of readout molecules} 
\cite{Mehta2012,govern2014,Govern_PRL_2014}.
 \teoedit{Other authors have highlighted the necessity of dissipation  in adaption
  \cite{Tu2012,Sartori2014,Ito2014} and kinetic proofreading
  \cite{Hopfield:1974uo,Murugan2012}.}

\teoedit{While it has been noted that there is a connection
  between the dissipation present in cellular copying and the thermodynamics of computation
  \cite{Mehta2012,govern2014,Govern_PRL_2014}, the nature of the connection remains 
nebulous. How do
    cellular protocols compare to the canonical copy
    protocols typically considered in the computational literature?
    Can cellular systems reach the fundamental thermodynamic limit on the
    accuracy and energetic cost of a measurement?  If not,  what is the underlying reason
  for the additional dissipation? Is it due to the nature of
  the biomolecular reactions, or due to the design of
    the signalling network? And if cellular systems cannot reach
    the fundamental limit, how does the trade-off between energy and
    precision differ from the ideal case?  To answer \teoresp{all of} these questions, it is necessary to
  construct a rigorous} mapping between cellular processes and
computational copying. Understanding the connection between the thermodynamics
of computation and the thermodynamics of biological processes
\cite{Tu2012, Endres,
  Mehta2012,Barato2013,Skoge:2013fq,Hopfield:1974uo,Ninio:1975vv,Qian2006,govern2014}
at a mechanistic level would enable translating quantitative, not
just qualitative, results from the literature on computation.

\teoresp{We rigorously map a canonical push-pull signalling motif to a
  computational copy device at the level of the master equation. We
  thereby identify a lower bound on dissipation that arises from the
  failure to exploit correlations generated between receptors and
  readouts (rather than the widely discussed costs of ``erasure''
 \cite{Landauer1961, lambson2011exploring, berut2012experimental, Jun2014,Hong2014,Mehta2012}). Our mapping demonstrates} that the push-pull network
cannot converge on this fundamental limit for a general input signal,
regardless of its parameters, and we prove that \teoeditnewb{more
  complex} biochemical networks \teoeditnewb{involving multiple steps
  or parallel pathways} cannot perform any better.  Remarkably,
however, cellular systems can operate close to the lower bound at
moderate-to-high accuracy, even at a high rate of copying. Further,
cellular networks are naturally adaptive, dissipating less when the
sampling challenge is reduced.  Finally we show that an artificial
copy device based on biochemical reactions can achieve the
thermodynamic bound with exogenous manipulation of fuel
concentrations, providing an alternative platform for investigating
the thermodynamics of computation.

\section{Push-pull system} 

\begin{figure}
\centering
\includegraphics[width=8.5cm]{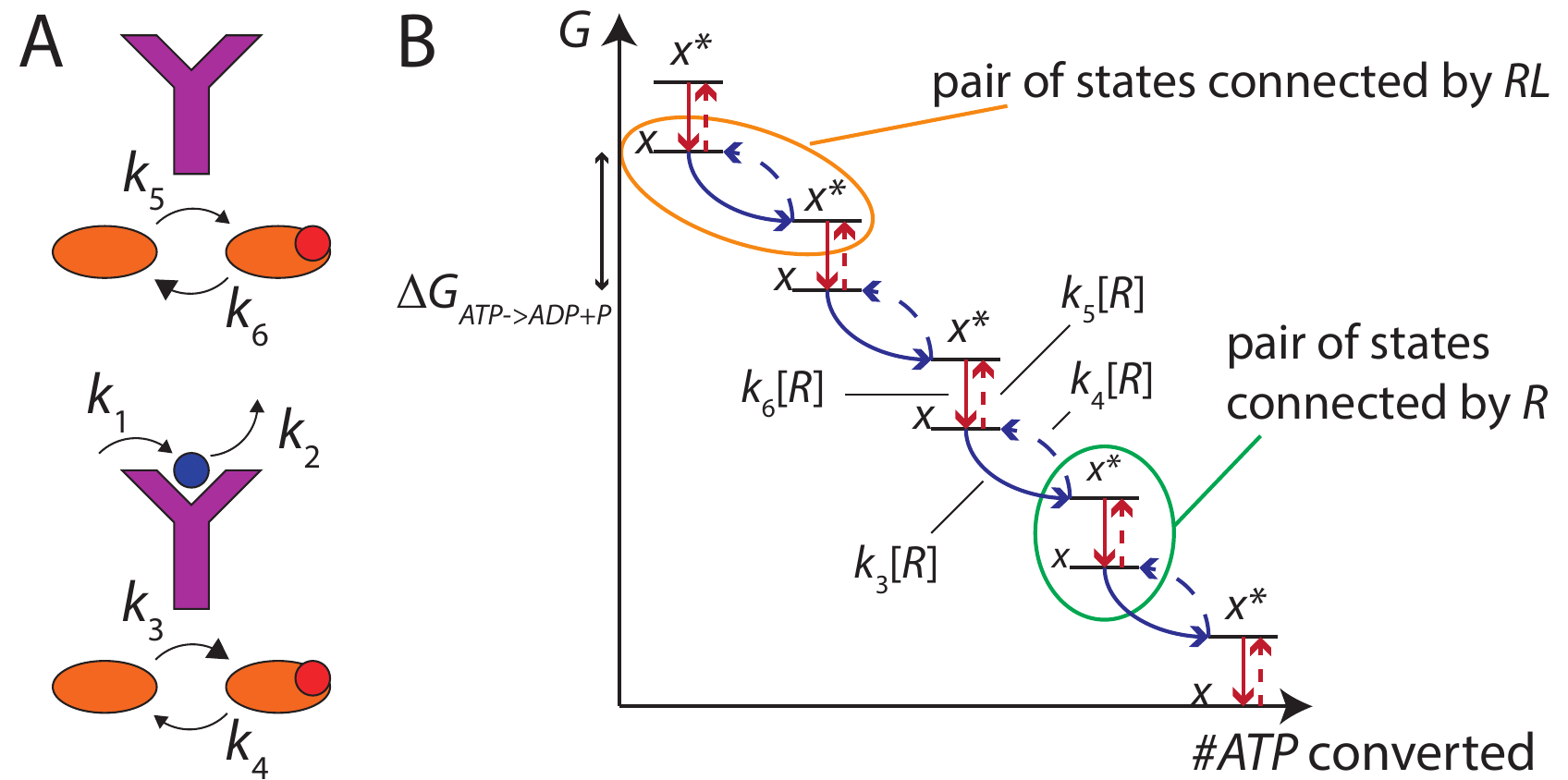}
\caption{A canonical signaling network. (A) The signaling
  network utilizes a receptor that acts as a bifunctional
  kinase/phosphatase: when bound to ligand, it catalyses the
  activation of the downstream readout; when unbound, it catalyzes the
  deactivation of the downstream readout.  Free-energy dissipation due
  to the use of fuel, coarse-grained from this representation, drives
  the reactions.  \teoedit{(B) Schematic free-energy
    landscape of a single readout molecule in the biochemical
    network. We plot the free energy $G$ as a function of the
    number of ATP molecules that are converted into ADP
      molecules, for the two states
    $x$ and $x^*$. Thermodynamically favourable transitions are shown
    with solid arrows, and unfavourable transitions with dashed
    arrows. The presence of catalysts $R$ and $RL$ makes these
    transitions faster, and thereby push the system towards $x$ or
    $x^*$, but the overall thermodynamic drive is fixed for both
    reactions.}
  \label{fig:system}}
\end{figure}

 \teoresp{To explore biochemical copying} {we consider bi-functional
  kinase systems, which are common in bacteria
\cite{Stock:2000ve}. The bi-functional kinase
  either tends to phosphorylate or dephosphorylate a readout $x$,
  depending on the ligand-binding state of the receptor to which the
  kinase is coupled (Fig. \ref{fig:system}). If the receptor $R$
    is bound to ligand $L$, then the bi-functional kinase acts as a
    kinase, catalyzing phosphorylation and dephosphorylation reactions
    that are both coupled to ATP hydrolysis. If the receptor is not
    bound to ligand, then the bi-functional kinase acts as a
    phosphatase, catalyzing phosphorylation and dephosphorylation
    reactions that are uncoupled from ATP hydrolysis. Such a system
    can be described by the following reactions}
\begin{equation}
\begin{array}{c}
R + L {\rightleftharpoons} RL,
\vspace{2mm}\\
 RL + x +{\rm ATP} {\rightleftharpoons} RL + x^* + {\rm ADP} ,
 \vspace{2mm}\\
 R + x^*  {\rightleftharpoons} R + x + {\rm P},
\end{array}
\label{reactions}
\end{equation}
\teoedit{where the kinase/phosphatase activity is coarse-grained into the
ligand-binding state of the receptor. Here $x$ and $x^*$ represent unphosphorylated and phosphorylated
readout states, respectively.}

 \teoeditnewb{For simplicity, we take $[L]$ to be constant, and assume the system is maintained in a non-equilibrium steady state: $[{\rm ADP}],\,[{\rm ATP}]$ and  $[{\rm P}]$ are fixed. We  treat phosphorylation and dephosphorylation as instantaneous second-order
reactions. Thus }
\begin{equation}
\begin{array}{c}
R \underset{k_2}{ \overset{k_1 [L]}{\rightleftharpoons}} RL,\,
 RL + x   \underset{k_4}{ \overset{k_3}{\rightleftharpoons}} RL + x^*,\,
 R + x^*  \underset{k_5}{ \overset{k_6}{\rightleftharpoons}} R + x .
\end{array}
\label{reaction rates}
\end{equation}
\teoedit{ The master equation for this system, and the chemical
  kinetics approximation, are provided in Supplementary Discussion 1. We emphasize that the 
  reactions within each equation are the
  microscopic reverses of each other, while the reactions of the second
  and third equations correspond to distinct reaction paths. This
  yields a thermodynamically consistent model.}

Qualitatively, this circuit performs copies (or measurements) in the following way.
 \teoqq{As spontaneous phosphorylation and dephosphorylation
  in the absence of the bi-functional kinase occur at a low rate}
which we take to be zero, the two chemical modification
states of the readout are analogous to two \teo{stable states of
  a memory} bit separated by a large barrier, as widely considered in the 
computational literature \cite{Bennett1982,Bennett2003,feynman1998feynman}. \teoedit{There are,
however, two separate paths between the two wells, via exchange of
phosphate with ATP and via exchange of phosphate with the cytosol. 
 Moreover, the ATP-independent 
  (de)phosphorylation reaction 
  has a high yield of $x$ in equilibrium, whereas the ATP-coupled reaction
 has an intrinsically high yield of $x^*$. The resultant extended free-energy landscape for a single readout that
explicitly considers ATP turnover is illustrated in Fig.\,\ref{fig:system}\,B.
 The presence of $RL$ lowers the barrier between pairs of states 
connected by ATP turnover, whereas the presence of $R$ lowers the barrier 
between states not connected by ATP turnover.
 In this way \teoresp{the receptor's binding state effectively 
 restricts the free-energy landscape to favour}
 either $x$ or $x^*$, which can then be thought of
  as copying the state of the receptor into the chemical modification
  state of the readout \cite{Mehta2012,govern2014}. In the next
    section we make 
this analogy concrete, allowing a quantitative analysis of biochemical copying.}

\teoedit{
Since  each readout molecule provides a stable
    memory of the receptor state, the readout
    molecules collectively \teoresp{provide information on the state of the receptor in the recent past}. This enables time integration
of the receptor signal and hence enhanced accuracy of concentration estimates.\cite{berg1977,Mehta2012,govern2014} We are not concerned, however,  with the precision of sensing a constant concentration \cite{berg1977,bialek2005,Mehta2012,Kaizu:2014eb,govern2014}, nor with the ``learning rate'' between the ligand concentration and the network, which is important when changes in external concentrations are more rapid  \cite{Barato2014, Hartich2014, Horowitz2014}. Rather,
    we are interested in whether the readout reaction
 can be rigorously described in terms of a copy
    process, and if so, how this cellular protocol compares to optimal quasistatic computational protocols involving the manipulation of energy landscapes
that are common in the literature\cite{Bennett1982,Bennett2003,feynman1998feynman}.
}


\section{Results}
\subsection{ The biochemical network as a copy process.}
\label{sec:rancopy}
\teoedit{
Let us consider the dynamics necessary for the biochemical network to \teoresp{be described rigorously as a stochastic copying system, in which randomly-selected data bits (receptors) are copied into randomly selected memory bits (readouts) at certain rates. To make such a mapping,}
each readout molecule should perform a copy of any one ligand-bound receptor at a rate $k_{RL}^{\rm copy}$, and any one unbound receptor at a rate $k_{R}^{\rm copy}$. These copies are performed with accuracies $s_{RL}$ and $s_{R}$ respectively, and the result of the copy \teoresp{should be} independent of the prior state of the readout. Thus a copy of $RL$ ($R$) returns $x^*$ ($x$) with probability $s_{RL}$ ($s_R$). \teoresp{Note that the state of the readout can be identical both before and after the copy, just as bits overwritten with new data may have the same value as before; transitions and copies are not equivalent. Indeed,} \teoresp{if the network is to be described as a copy process, then} readouts in state $x$ should be converted into readouts  in state $x^*$ at a rate 
\begin{equation}
\sigma_{x \rightarrow x^*} = (N_{RL} k_{RL}^{\rm copy}  s_{RL} + N_R k_{R}^{\rm copy }  (1-s_{R}) ) N_x,
\label{eq:copy1}
\end{equation}
and $x^*$ should be converted into $x$ at a rate
\begin{equation}
\sigma_{x^* \rightarrow x} = (N_{RL} k_{RL}^{\rm copy} (1-s_{RL}) +N_ R k_{R}^{\rm copy}   s_{R} ) N_{x^*}.
\label{eq:copy2}
\end{equation}
Here, $N_R$ and $N_{RL}$ are numbers of receptors, and \teoresp{$N_x$ and $N_{x^*}$ the numbers of readouts,
in each state.}
\teoresp{By definition, copies are made at a  rate $(k_{RL}^{\rm copy} N_{RL} + k^R_{\rm copy} N_R)  N_{x_T} $, with $N_{x_T}$ the total number of readouts.}

Returning to our actual  biochemical network, Eq. \ref{reaction rates} specifies the dynamics of $x$ and $x^*$.  The transitions occur with rates
\begin{equation}
\begin{array}{c}
\sigma_{x \rightarrow x^*} = (k_3 N_{RL} +  k_5 N_{R} ) [x], \\
\\
\sigma_{x^* \rightarrow x} = (k_4 N_{RL} +  k_6 N_{ R} ) [{x^*}].
\end{array}
\label{eq:copy3}
\end{equation}
Comparing Eqs.\ref{eq:copy1}-\ref{eq:copy3} , we see that the biochemical network and stochastic copy process are equivalent at the level of transition rates, which specify the master equation, if
}
\begin{equation}
k_{RL}^{\rm copy}  = (k_3 + k_4)/V ,\,\, k_{R}^{\rm copy}  = (k_5 + k_6)/V,
\end{equation}
in which $V$ is the volume of the system and the copy accuracies are
\begin{equation}
s_{RL} = {k_3}/{(k_3 + k_4)},\,\, s_R= {k_6}/{(k_5 + k_6)}.
\label{s1 and s2}
\end{equation}
Thus the average copy rate is
 \begin{equation}
\dot n_{\rm copy} =  [{x_T}] \left( (k_3+k_4)\langle N_{RL} \rangle  + (k_6+k_5)\langle N_{R} \rangle \right),
\label{n dot copy}
\end{equation}
in which square brackets indicate a concentration within the volume $V$.
The biochemical network can therefore be directly mapped to a stochastic copying process.

\subsection{ Energetic cost per copy cycle.}
In the non-equilibrium steady state, \teoedit{the average dissipation rate of chemical free energy   ($\dot{w}_{\rm chem}$) is} 
\begin{equation}
\dot{w}_{\rm chem} = - \dot n_{\rm flux} (\Delta \mu_1 + \Delta \mu_2) =  \dot n_{\rm flux} kT \ln \left( \frac{k_3 k_6}{k_4 k_5} \right),
\label{DGdot}
\end{equation}
where $\dot n_{\rm flux}$ is the \teoedit{average} current of readout molecules around the phosphorylation/dephospohorylation loop; $\Delta \mu_1 = \mu_{\rm ADP} - \mu_{\rm ATP}$ and  $\Delta \mu_2 = \mu_{\rm P}$. The sum $-(\Delta \mu_1 + \Delta \mu_2)$ is the free energy of ATP hydrolysis, which is dissipated when a readout goes around this cycle once.

 \teoedit{To proceed, we introduce the following averages:
\begin{equation}
p = \frac{\langle N_{RL} \rangle}{N_{R_T}} =  \frac{k_1 [L] } {k_2 + k_1[L]}, \hspace{2mm}
f = \frac{\langle [x^\ast] \rangle }{ [x_T]},
\end{equation}
with $N_{R_T} = N_{RL} + N_{R}$ the total number of receptors. In the mean-field limit, $\dot n_{\rm flux}$ follows from Eq. \ref{reaction rates} as 
\begin{equation}
\dot n_{\rm flux} =\langle  k_3 N_{RL} [x] - k_4 N_{RL} [x^*] \rangle = (k_3 (1-f) - k_4 f) p [x_T] N_{R_T}.
\label{n dot flux}
\end{equation}
 Furthermore, the fraction of phosphorylated readout $f$ is
\begin{equation}
f = \frac{k_3 p +k_5(1-p)}{(k_3+k_4)p + (k_5 + k_6)(1-p)},
\label{f-equation}
\end{equation} 
giving
\begin{equation}
\dot{w}_{\rm chem} = \frac{(k_3 k_6 - k_4 k_5) p(1-p) [x_T] N_{R_T}}{(k_3+k_4)p + (k_5 + k_6)(1-p)} kT \ln \left( \frac{k_3 k_6}{k_4 k_5} \right).
\label{G dot}
\end{equation}
For the full calculation, we refer to Supplementary Discussion 1. The mean field approach  holds in the limit of many receptors,
or when the \teoresp{readout phosphorylation dynamics is slower than the receptor-ligand dynamics, as required for the mechanism of time integration}. If these conditions are not met, \teoresp{a given readout performs many copies of autocorrelated data}. Detailed analysis of this regime is beyond the scope of this
work, but a brief discussion is provided in Supplementary Discussion 1.   }

Given the rate of copying (Eq.\,\ref{n dot copy}) and the rate at which chemical work is done (Eq.\,\ref{G dot}), we are now in a position to calculate 
the  chemical work done per copy \teoedit{cycle}:
\begin{equation}
\frac{w_{\rm chem}}{n_{\rm copy}} = \frac{(k_3 k_6 - k_4 k_5) p(1-p)}{\left((k_3+k_4)p + (k_5 + k_6)(1-p)\right)^2} kT \ln \left( \frac{k_3 k_6}{k_4 k_5} \right).
\label{work per copy}
\end{equation}
This result can be simplified by noting that the probability a copy is made of $RL$ is not $p$, but rather
\begin{equation}
 p^\prime = p\frac{k_3+ k_4}{(k_3+ k_4) p + (k_5 + k_6) (1-p)}.
 \label{pprime}
\end{equation}
Indeed,  if $k_3+k_4 > k_5+k_6$,  a given ligand-bound receptor is more
frequently copied than a given  unbound receptor molecule (see
  Fig. \ref{fig:system}). \teoresp{Using the expression for
  $p^\prime$, Eq.\,\ref{pprime}, the fractional yield of phosphorylated readout can be written in the intuitive form $f = p^\prime s_{RL} + (1-p^\prime)s_R$, and Eq.\,\ref{work per copy} can be simplified to}
\begin{equation}
\frac{w_{\rm chem}}{n_{\rm copy}}  = {(s_R+s_{RL}-1) p^\prime (1-p^\prime)} kT\ln \left( \frac{k_3 k_6}{k_4 k_5} \right).
\label{eq:w_chem_biochemical0}
\end{equation}
The quantity $ kT\ln \frac{k_3 k_6}{k_4 k_5}$ has dimensions of energy and is equal to the chemical work done during a single phosphorylation/dephosphorylation cycle. We can split it into an energy related to the accuracy of copying $RL$, $E_{s_{RL}} = kT \ln (k_3/k_4)= kT \ln(s_{RL}/(1-s_{RL}))$, and an energy related to the accuracy of copying $R$, $E_{s_R} = kT \ln (k_6/k_5)=kT \ln(s_{R}/(1-s_{R}))$.
\begin{equation}
\frac{w_{\rm chem}}{n_{\rm copy}}  = {(s_R+s_{RL}-1) p^\prime (1-p^\prime)} (E_{s_R} + E_{s_{RL}}).
\label{eq:w_chem_biochemical}
\end{equation}
 \teoresp{Note that, although $E_{s_R}+E_{s_{RL}} = -\Delta \mu_1 - \Delta \mu_2$, $E_{s_R}$ and $E_{s_{RL}}$ incorporate differences in internal free energy between $x$ and $x^*$ and hence $E_{s_{RL}} \neq -\Delta \mu_1$ and $E_{s_{R}} \neq -\Delta \mu_2$ in general.}
 \teoqqq{Inverting the sign of $E_{s_R}$ and $E_{s_{RL}}$ corresponds to a mirror-image encoding, in which $RL$ is copied to $x$ and $R$ to $x^*$. Low accuracy  corresponds to $E_{s_R}, E_{s_{RL}} \rightarrow 0$, when  $s_R \approx \frac{1}{2} (1 + E_{s_R}/2kT)$, and  $s_{RL} \approx \frac{1}{2} (1 + E_{s_{RL}}/2kT)$.  In the symmetric case of $ E_{s_R}= E_{s_{RL}}= -\Delta \mu/2$, the accuracy of copies is $s_{R} = s_{RL} =  \frac{1}{2} (1 -\Delta \mu /4kT)$. A similar analysis, with an equivalent outcome, is performed in Supplementary Discussion 2 for a related system in which receptors act only as kinases.}

\teoresp{
\subsection{ Optimal devices set a thermodynamic bound on the dissipation of the biochemical network.}}
The  biochemical network has the dynamics of a process in which readouts randomly perform copies of receptors. It is also fundamentally dissipative -- does
\teoeditnewb{the fact that it is a copy process set a practically relevant thermodynamic bound on this dissipation?}

We can calculate the minimal thermodynamic cost of the random computational copy process that is equivalent to our biochemical network motif. In each copy operation \teoresp{of this equivalent process,  a memory bit $M$ is exposed to an initially uncorrelated data bit $D$ (in state 1 with probability $p(d=1) = p^\prime$ as defined in Eq.\,\ref{pprime}) - the final result is a memory bit in state $1$ with probability $p(m=1|d=1) = s_{RL}$ if $D$ is in state $1$, or probability  $p(m=1|d=0) = 1-s_{R}$ if $D$ is in state $0$, giving a marginalised probability of $p(m=1)=f= p^\prime s_{RL} + (1-p^\prime)s_R$. } Mutual information $I$ is then generated between the data and the memory \teoresp{during each individual copy of the equivalent process, with
\begin{equation}
 \begin{array}{c}
  I(s_R, s_{RL}, p^\prime) = 
 p^\prime s_{RL} \ln \left(\frac{s_{RL}}{f} \right) + p^\prime(1-s_{RL}) \ln \left( \frac {1-s_{RL}}{1-f} \right) \vspace{2mm} \\
+ (1-p^\prime) s_R \ln \left(\frac{s_R}{1-f} \right) + (1-p^\prime)((1-s_R) \ln \left( \frac {1-s_R}{f} \right).
\end{array}
\label{eq:info}
\end{equation}
We emphasize that the above expression is simply the mutual information between  bits $M$ and $D$ at the end of a single discrete copy operation, calculated directly as $I(M,D) = \sum_{m,d} p(m,d) \ln (p(m,d)/p(m)p(d))$. Here $p(m,d)$ is determined straight-forwardly from $p(m,d) = p(m|d) p(d)$, using the expressions above for $p(d)$ and $p(m|d)$ in terms of $p^\prime$ and the measurement accuracies $s_R$ and $s_{RL}$. }

\teoresp{Immediately after copying}, the data (receptor) and memory (readout) are decoupled without loss of information. This means that they remain correlated, even though there is no direct physical interaction between the data and memory anymore. 
Generating correlations that persist after direct interactions cease
implies pushing the combined system out of equilibrium: the free
energy required is $kTI$ \cite{Sagawa2009, Horowitz2013,
  Parrondo2015}. Making computational copies thus amounts to storing
free energy in mutual information, and if this information is not used
to extract work (as done by an efficient Maxwell Demon) but simply
lost in an uncontrolled fashion  then the process is irreversible and
the information lost sets a lower bound on dissipation. In the random
computational  copy process to which we map our biochemical network (see section \ref{sec:rancopy}), the stored information is not used to extract work, and hence $kTI$ sets a lower bound on entropy generation for the entire cyclic copy operation \cite{Sagawa2009}. \teoresp{For completeness,} a typical copying device and a set of quasistatic protocols that can achieve this bound for various input data are given in Supplementary Fig. 1 and analysed in Supplementary Discussion 3.

\begin{figure*}
\centering
\includegraphics[width=3.8cm, angle=-90]{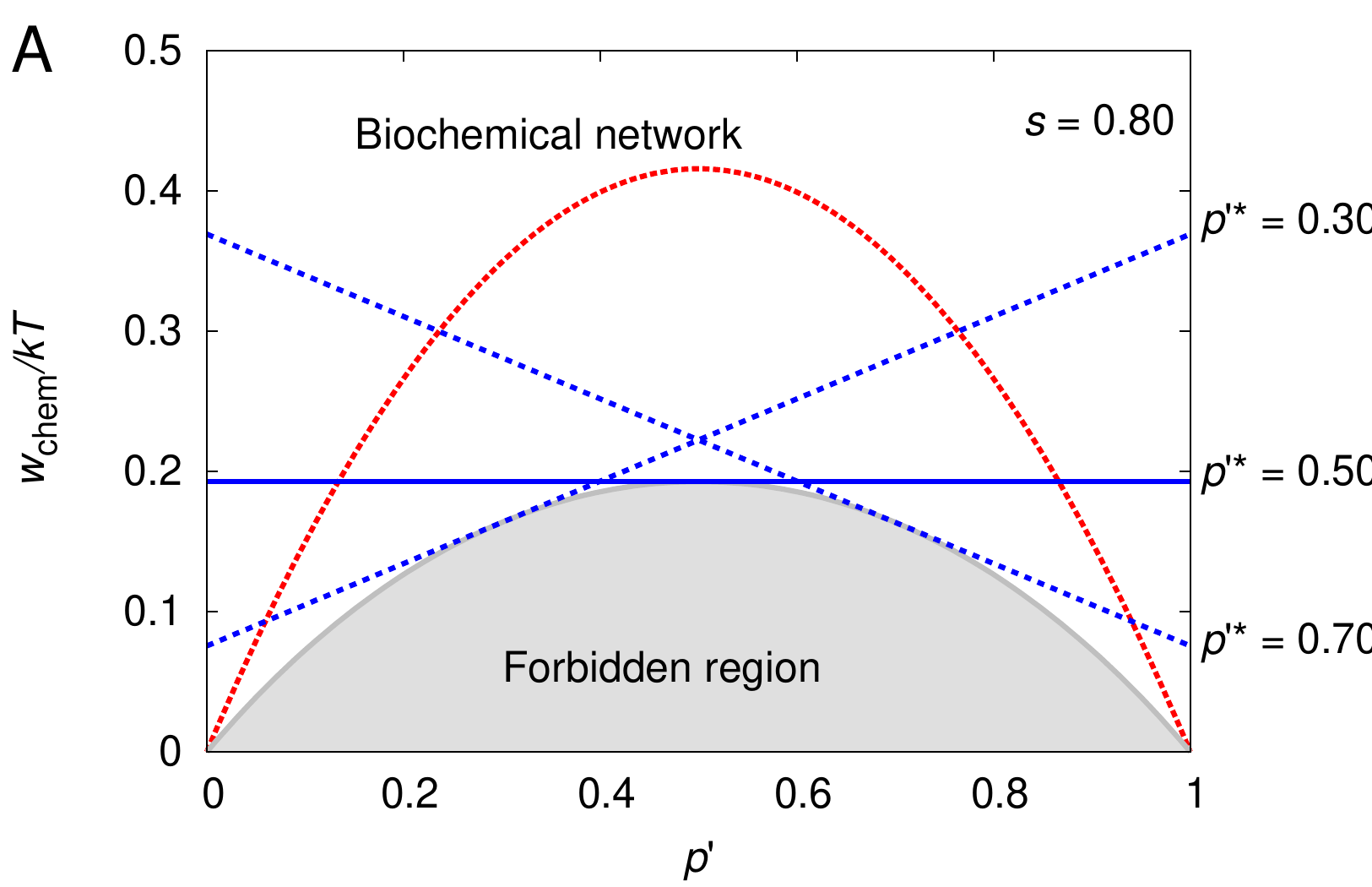}
\includegraphics[width=3.8cm, angle=-90]{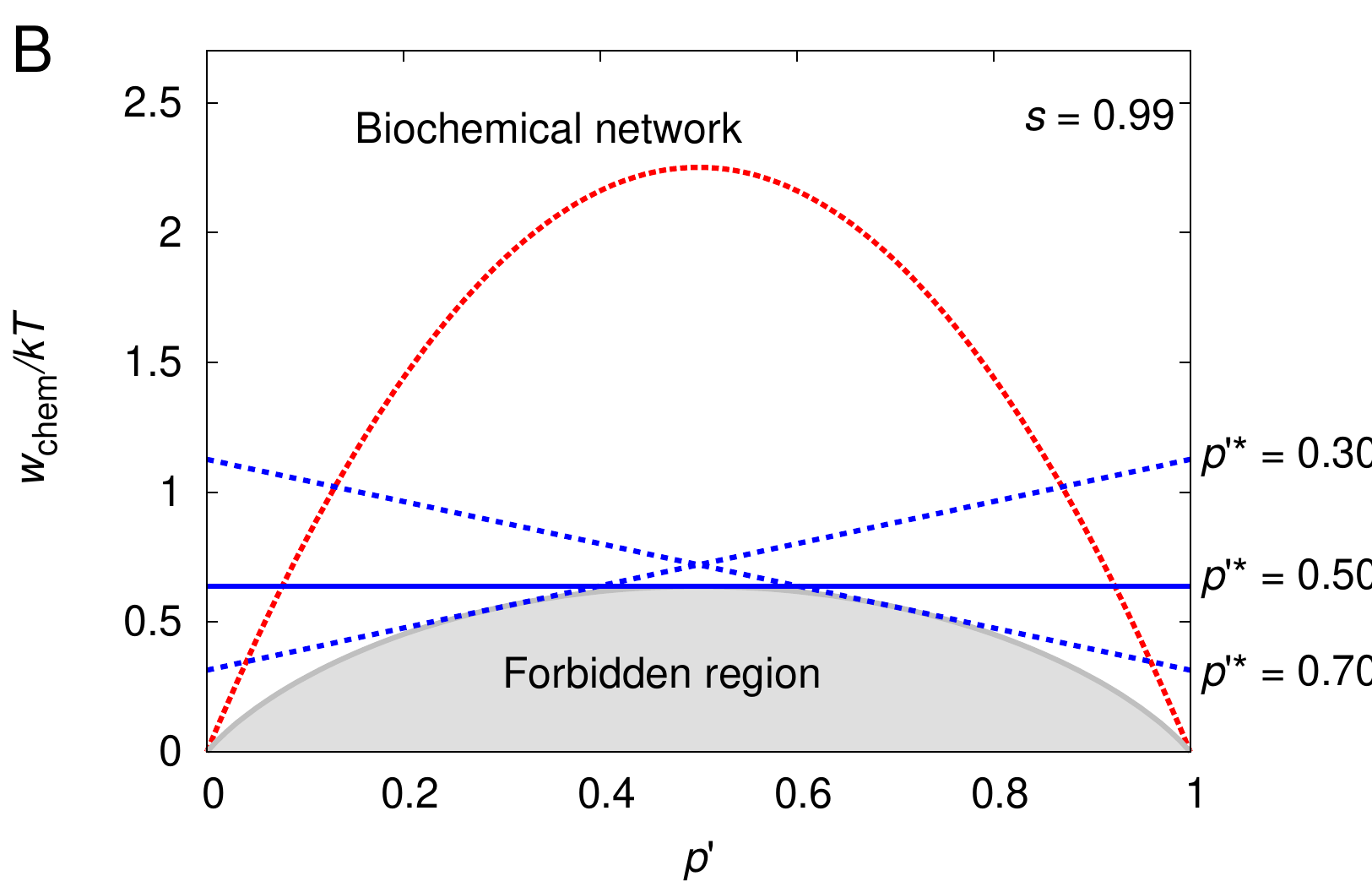}
\includegraphics[width=3.8cm, angle=-90]{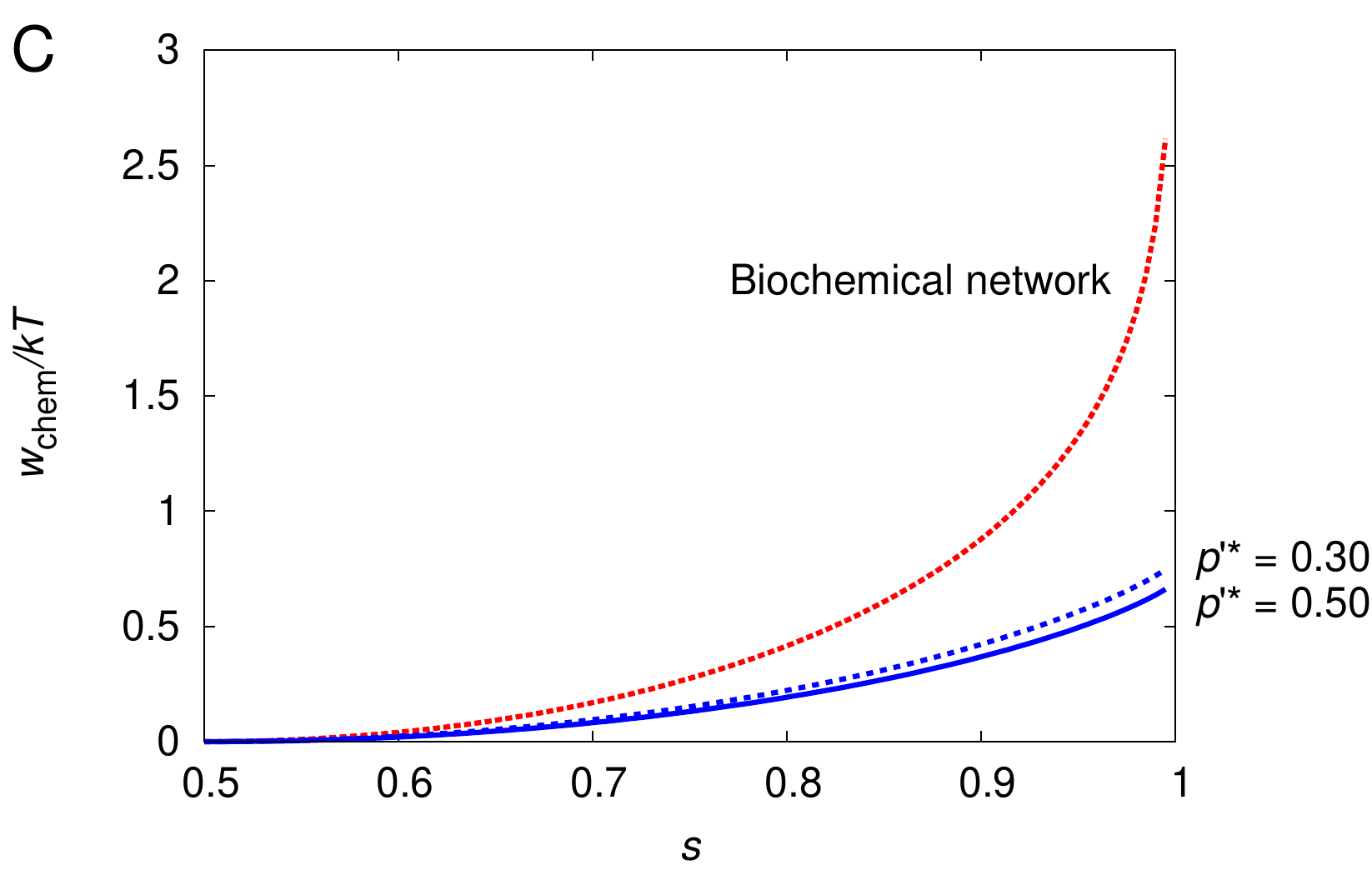}
\caption{Trade-off between dissipation and accuracy in \teoedit{copying}.  \teoqq{The (chemical) work per copy \teoedit{cycle}} for different probabilities $p^\prime$ of attempting to copy $RL$ is plotted at two different values of the measurement accuracy: (A) $s=0.80$; and (B) $s = 0.99$. \teoqq{Note that $p^\prime=p$, the probability that a receptor is in state $RL$, if $k_3+k_4$ = $k_5+ k_6$ when the sampling rate for the two receptor states is the same.}  The biochemical implementation (red line) does not achieve the lower bound for a measuring device, which is the border of the shaded region. The blue lines correspond to quasistatic protocols  for the device in Supplementary Fig. 1 (see Supplementary Discussion 3) and are optimal for a specific \teoqq{value of $p^\prime$}, $p^{\prime *}$, at the given values of $s$. 
(C) Dissipation per copy \teoedit{cycle} at $p^\prime=0.5$ as a function of $s$ for the biochemical network (dashed red line) and a system that saturates the bound (solid blue line). The solid blue line ($p^{\prime *}=0.5$) saturates at $kT \ln 2$ for perfect accuracy ($s=1$) and the cost of the canonical biochemical motif diverges. \label{fig:tradeoff}  }
\end{figure*}

\begin{figure}
\centering
\includegraphics[width=8cm]{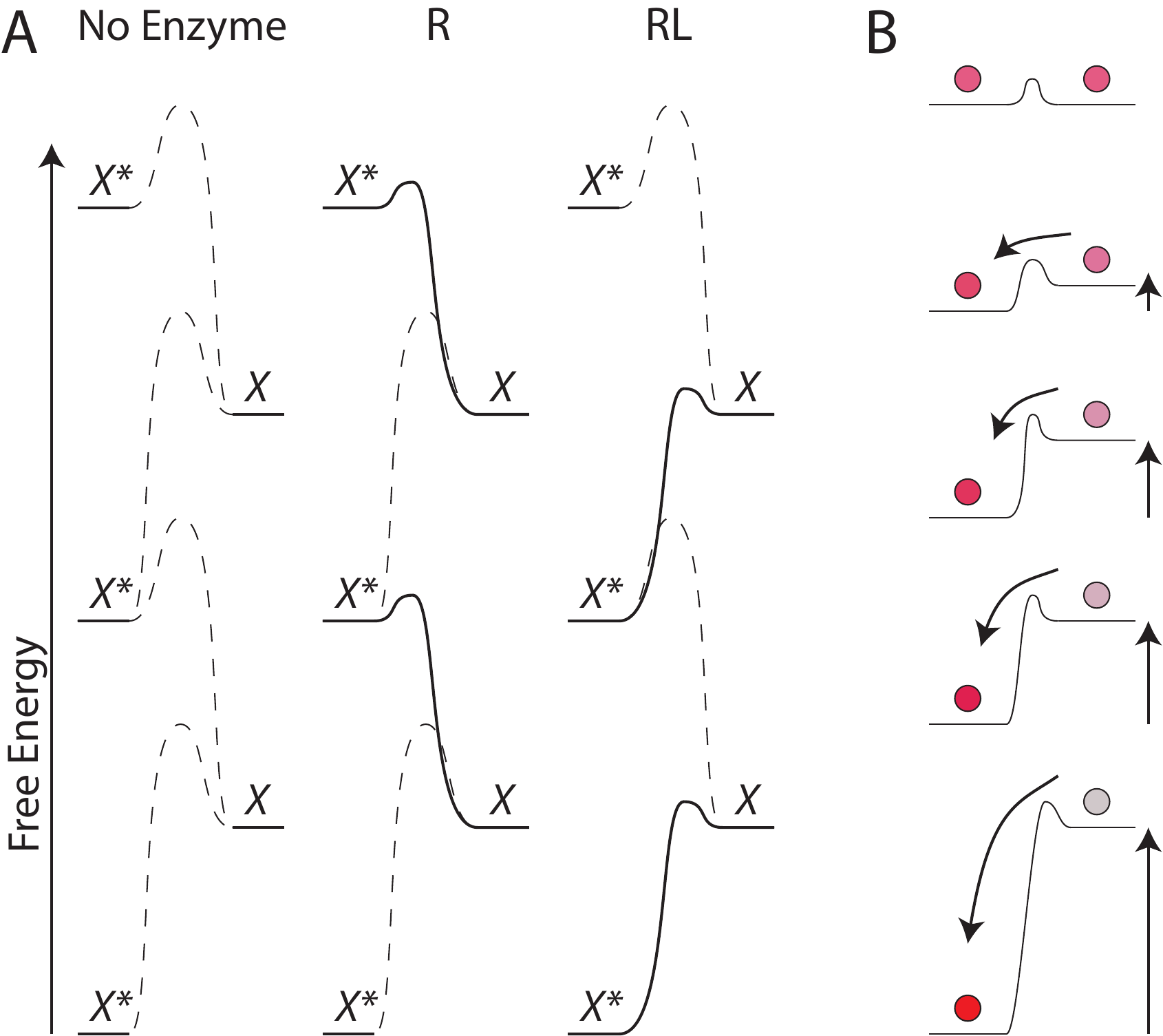}
\caption{Schematic comparison between the (free-) energy landscapes of the biochemical network (A) and a thermodynamically optimal protocol (B). In (A), a ladder of states  exists, with successive rungs related by the turnover of a single ATP (implicit in this figure, explicit in Fig.\,\ref{fig:system}). In the absence of an enzyme, all transitions are slow (indicated by the dashed free-energy barrier connecting the states). In the presence of $R$, half of the transitions are catalysed and can occur rapidly (solid lines); when exposed to $RL$, the alternative transitions are rapid. The heights of the rungs, however, are fixed, meaning that
all transitions involve a fixed amount of chemical work equal to the offset. By contrast, for the optimal protocol  in (B), switching between states is driven by slowly destabilizing one state with respect to the other, so that the majority of transitions have already occured before the offset approaches its limiting value.
\label{fig:landscapes}}
\end{figure}

\teoresp{The biochemical network is equivalent to  the random copy process that generates information $I(s_R, s_{RL}, p^\prime)$ given by Eq.\,\ref{eq:info} at each copy event, and work is not extracted from the information generated. Its entropy generation is thus constrained by the informational bound: }
\begin{equation}
\frac{w_{\rm chem}}{kT n_{\rm copy}}  \geq  I(s_R, s_{RL}, p^\prime).
\label{bound}
\end{equation}
This bound implies an efficiency $\eta ={(k  T I )}/ {(w_{\rm chem}/n_{\rm copy})} \leq 1$. 

It is important that the readout molecule states are persistent -- the ligand-binding state of the receptor is remembered even after detaching, thus enabling time integration \cite{govern2014,Govern_PRL_2014}. Systems that do not make persistent copies, including passive readouts \cite{Govern_PRL_2014}, the receptors themselves, \teoresp{the formation of templated copolymers in which the copy remains bound to the template \cite{Bennett1979, Andrieux2008,Sartori2013}}, and also some more complicated \teoresp{sensing} networks \cite{Sartori2014}, are not bound by equivalent limits. Previous work has linked energy dissipation in similar systems to the erasure of the memory\cite{Mehta2012,govern2014}, that is the resetting of the memory bit to a well-defined state. As shown by Landauer, this erasure does lead to the transfer of heat from the system to the surroundings. Resetting a bit, however, is not intrinsically thermodynamically irreversible\cite{Bennett1982,Bennett2003,Maroney2005,Fahn1996,Granger2013} -- indeed, Landauer's calculation of the minimal heat transfer applies to a thermodynamically reversible erasure step. It is therefore \teoresp{the failure to extract work from correlations} \cite{Fahn1996}, rather than erasure itself, which is the origin of the thermodynamic irreversibility of a copy cycle. It is also this failure that sets the fundamental thermodynamic lower bound on the work to perform copy cycles. In fact,  erasure is not even necessary in an optimal copy cycle, which leaves open the possibility that the biochemical network, which actually contains no explicit erasure, could achieve the lower bound of Eq.\,\ref{bound}.

\vspace{5mm}

\subsection{Trade-off between  dissipation and error.}
The bound in Eq.\,\ref{bound}  holds for all choices of $s_{RL}$, $s_R$ and $p^\prime$. But how close does the biochemical network come to this bound, and can  it  be reached in certain conditions? Our concrete mapping allows us to examine these questions. Initially, we consider the simplest case of equal accuracy ($s_R = s_{RL}=s$). The dissipation per copy \teoedit{cycle} for the biochemical network (Eq. \ref{eq:w_chem_biochemical}) then reduces to
\begin{equation}
\frac{w_{\rm chem}}{n_{\rm copy}} = 2 E_s  (2s -1) p^\prime(1-p^\prime),
\label{w_simple}
\end{equation}
with $E_s = E_{s_{RL}}= E_{s_R}=-(\Delta \mu_1 + \Delta \mu_2)/2$,  half of the free energy released by the breakdown of ATP into ADP and P. The chemical work per copy \teoedit{cycle}, ${w_{\rm chem}}/{n_{\rm copy} }$, is plotted against $p^\prime$ in  Fig. \ref{fig:tradeoff}\,A-B for two values of $s$ (red dashed lines). We also indicate behaviour forbidden by the thermodynamic bound of Eq.\,\ref{bound} (grey region).  In Fig. \ref{fig:tradeoff}\,(C), we fix $p^\prime = 0.5$ and plot the cost per copy cycle against $s$ for the biochemical network (red dashed line) and an optimal system that saturates the bound (solid blue line).

Both the cost of biochemcial network and the thermodynamic bound drop  as the required accuracy is decreased. The irreversible loss of information bounds the  work, and lower accuracy implies less information to be lost. 
It is clear from Fig. \ref{fig:tradeoff}\,(C), however, that the two systems have a very different tradeoff between
dissipation and accuracy. The required free-energy input for the
biochemical case diverges as $s \rightarrow 1$ \teoqqq{(${w_{\rm
      chem}}/{n_{\rm copy}} \approx - 2 p^\prime(1-p^\prime)
  kT\ln(1-s)$)}, whereas the dissipated work remains finite in an
optimal system. It is clear that a cell will have to sacrifice some accuracy for the sake of efficiency.
Even in the limit of low accuracy ($s \rightarrow \frac{1}{2}$) and
  at $p^\prime=0.5$,   the biochemical network is  twice as costly as the  bound ($\eta \rightarrow \frac{1}{2}$ from above).  Expanding Eqs.\,\ref{bound} and \ref{w_simple}  for $s \rightarrow \frac{1}{2}$ gives 
  ${w_{\rm chem}}/{n_{\rm copy}} \approx 4 kT (s-\frac{1}{2})^2$ for the
biochemical network and ${w_{\rm chem}}/{n_{\rm copy}} \approx 2 kT
  (s-\frac{1}{2})^2$ for the optimal device and protocol. 

\teoeditnewb{The fundamental difference between the biochemical
  network and an optimal protocol is illustrated in
  Fig.\,\ref{fig:landscapes}. An optimal protocol requires reversible
  quasistatic manipulation of \teoresp{(free)}-energy levels
  \teoresp{over time} \cite{Bennett1982,Bennett2003}.  In the biochemical network, however, the overall differences in \teoresp{free-energy} between  levels are fixed \teoresp{over time, and not slowly varied during a measurement.  Instead, the receptors selectively catalyse specific reactions dependent on the receptor state.  Because the free-energy levels, and hence the energetic drive for the copy process, are constant, the cellular copy protocol is energetically more costly than the thermodynamically optimal one.}}

\teoeditnewb{This difference is particularly intuitive in the limits of low and high copy accuracy for $p^\prime =0.5$.}
 For the biochemical network, the difference
  between the fraction of correct copies $s$, and the fraction of
  incorrect copies $1-s$, is $2(s-\frac{1}{2})$. For $ p^\prime =0.5$,
 in only half of these
  cases does a reaction  occur and so the net number of
  reactions in the intended direction is $s-\frac{1}{2}$ per copy
  \teoedit{cycle}. Each of these net reactions has a cost given by the
  driving free energy, $E_s \approx 4(s-\frac{1}{2})kT$, giving an overall
  cost per copy \teoedit{cycle} of ${w_{\rm chem}}/{n_{\rm copy}}
  \approx 4 kT (s-\frac{1}{2})^2$. For an optimal quasistatic protocol, a similar
  argument can be constructed (see Supplementary Discussion 3), but the average energy offset
between the states  at which a transition occurs is less than the final offset required by the accuracy, $E_s$.
In the limit of $E_s \rightarrow 0$, we obtain an average cost  of $E_s/2$, rather than $E_s$ as for the biochemical
  network.  


\teoedit{In the limit of
    high accuracy, $s \to 1$ and $E_s \rightarrow \infty$, a similar analysis shows why the biochemical network is much less efficient than an optimal quasistatic protocol. The cost of
    each transition continues to grow linearly with $E_s$  in the biochemical network, explaining the divergence of chemical work with
    $s$.
 In an optimal protocol, however, the work saturates when $E_s \gg kT$. In this case, the energy difference between the two states is raised quasistatically to $E_s$. Therefore  at each moment in time,
      the bit is in equilibrium. As $E$ is
      raised, the probability that the bit is in the high energy state
      decreases, until it rapidly becomes negligible when $E > k T$: from hereon
      the higher energy state can be raised further without any additional cost.}

\teoeditnewb{As is evident from Figs.\,\ref{fig:tradeoff}\,(A,B), the cost of both an optimal system and the biochemical network decrease as we move away from $p^\prime = 0.5$ at fixed $s$. Fixed accuracy measurement results in less information if the data itself has low entropy, explaining the reduction in the bound. For the biochemical network, if the readout is exposed more often to one receptor state, it is more likely to be in the appropriate output state prior to a copy. Hence, fewer transitions are needed and less dissipation occurs.  For an alternative network in which receptors only function as kinases, this automatic compensation only occurs at low $p$ (see Supplementary Discussion 2).

\teoresp{The adaption to low entropy data is so effective that}
$\eta$  is actually highest for $p^\prime \rightarrow 0$ or 1 (when $\eta \rightarrow \frac{1}{2}$ from below for all $s$), and lowest at $p^\prime =\frac{1}{2}$ (shown explicitly in Supplementary Fig. 2). This intrinsic adaption is fundamentally different from the behaviour of typical copying architectures.\cite{Bennett1982,Bennett2003}  For these devices the protocol must be parametrically adjusted for optimality as $p^\prime$ is varied. Figs.\,\ref{fig:tradeoff}\,(A,B) show, with blue lines, the work per copy of three fixed protocols that are optimal for $p^{\prime *} =0.3$, 0.5 and 0.7, respectively; each line is tangent to the forbidden region at the specific value of $p^\prime = p^{\prime *}$, but above it elsewhere, as derived in Supplementary Discussion 3.}

\teoresp{
While the different ``optimal protocols'' corresponding to the blue dashed lines in Fig.\,\ref{fig:tradeoff} are each optimal for a given $ p^{\prime *}$, they do not perform better than the biochemical network for all values of $p^\prime$. Moreover, the details of a cycle depend on $p^{\prime *}$, and so implementing an efficient protocol with $p^{\prime *} \approx p^\prime$ would require an estimate of $p^\prime$. In our cellular context, however, $p^\prime$  is precisely the} quantity that the system is trying to measure (rather than the state of individual receptors given a known $p^\prime$). The best that could be done, therefore, would be to pick a particular protocol, and update it as more information became available. Clearly, however, implementing this behaviour in an autonomous device would require substantial additional complexity. 

We have shown that the biochemical network cannot reach the fundamental
  limit of efficiency for $s_R= s_{RL}$, regardless of the system parameters. 
Neither reducing the reaction rates nor the thermodynamic driving  permits $\eta > {1}/{2}$. It is sometimes assumed that systems under steady-state forcing are effectively quasistatic or reversible in the limit of this forcing being weak \cite{Mehta2012} -- clearly that
is not the case here. For $s_R\neq s_{RL}$, it is possible to obtain an efficiency $\eta > {1}/{2}$, and $\eta$ can even approach unity for extreme values of $p^\prime$, $s_R$ and $s_{RL}$ (see Supplementary Fig. 3 and Supplementary Discussion 4). Nonetheless, it remains true that $\eta$  cannot converge on unity for general $p^\prime$, regardless of how $s_R$ and $s_{RL}$ are varied. The biochemical network is more dissipative than an optimal process, because operating at a constant thermodynamic driving force leads to more energetically expensive transitions than slowly manipulating energy levels. 

Thus far we have emphasized differences between the thermodynamic bound and the cost of the biochemical protocol. It is remarkable, however,
that  the biochemical network comes so close to the lower bound, even at fairly high accuracies. Reaching 99\% 
accuracy for $p^\prime=0.5$ requires less than four times the dissipation of the lower bound, and $\eta$ is even higher for $p^\prime \neq 0.5$. Further, this efficiency can be achieved at an arbitrarily high rate of copying  -- the absolute rates do not enter the expression in Eq.\,\ref{w_simple}. Through our quantitative mapping, we have  shown that a physically reasonable model system operating autonomously at an arbitrary rate and with a high copying accuracy comes close to the fundamental thermodynamic limit on the cost of a copy process.

\subsection{One-step copy processes are maximally efficient for the biochemical network.}
\begin{figure}
\centering
\includegraphics[width=8.5cm]{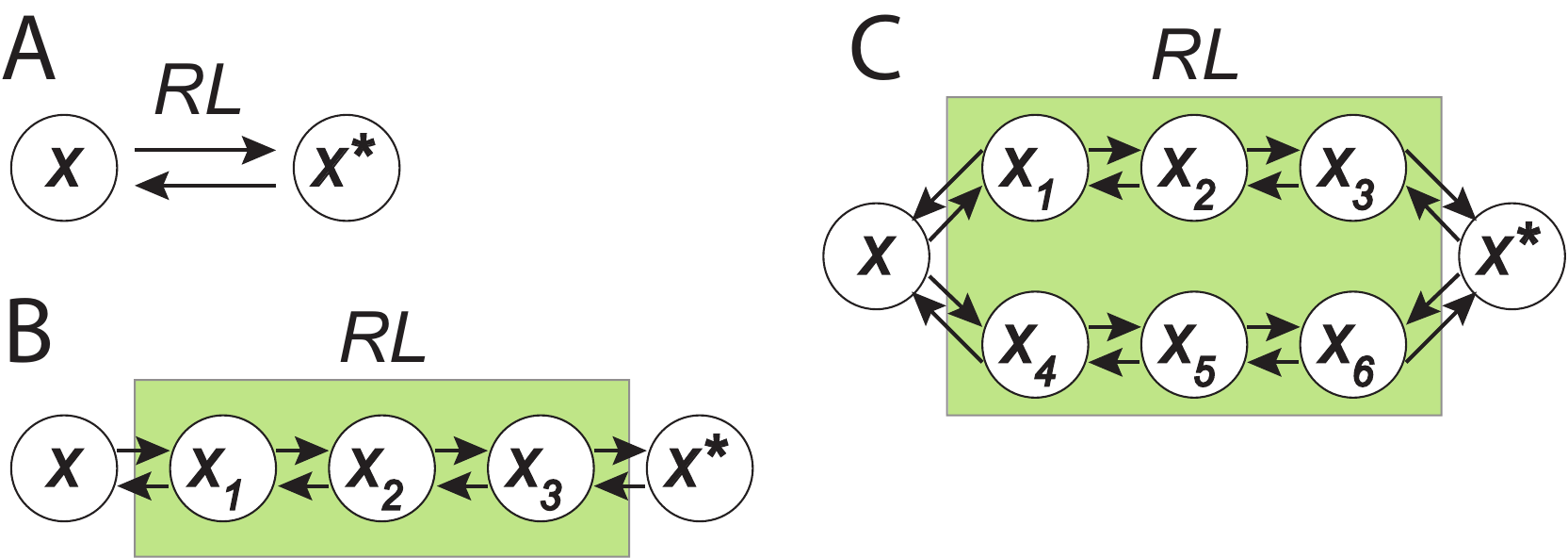}
\caption{ Copy processes of increasing complexity. (A) A simple one-step process in which the $RL$-bound state is assumed short-lived. (B) A tightly-coupled process in which the readout passes through multiple states whilst bound to the receptor, but only one pathway is possible. (C) A more general process in which multiple pathways between $x$ and $x^*$ exist. Each different path {\it could} be associated with different entropy increases in the environment.
\label{pathways}}
\end{figure}

The push-pull network considered so far is a one-step copy process, with the conversion between the phosphorylation states occurring via a single instantaneous transition.  \teoeditnewb{We now show that more complex processes, involving multiple steps or pathways, cannot improve the trade-off between dissipation and precision in autonomous, steady-state systems driven directly by an out-of equilibrium chemical fuel. We again consider  Markov process with discrete states, but we now allow for multiple states and multiple parallel pathways between $x$ and $x^*$ (see Fig.\,\ref{pathways}). We explicitly consider the interconversion of $x$ and $x^*$ by $RL$; an equivalent argument holds for reactions mediated by $R$. The autonomous requirement prohibits external control and implies that transition rates are fixed \teoresp{over time}.}

 In the simple one-step process, the accuracy of copying ${RL}$ is
 $s_{RL}= k_3/(k_3+k_4)$. In a more general process, however,
 transitions between $x$ and $x^*$ are not instantaneous and hence
 cannot be described with rate constants. The natural generalisation
 is to consider the flux $\phi$ from $x$ to $x^*$ and vice versa,
 which is defined as the rate at which trajectories leave $x$ and
 subsequently reach $x^*$ instead of returning to $x$
 \cite{Allen2009}.  The fluxes determine the copy accuracy in the same way  as the rate constants for the one-step process, since they represent the rate receptors initiate correct and incorrect transitions between $x$ and $x^*$.

Consider the accuracy of copying $RL$, $s_{RL}= \phi_3/(\phi_3+\phi_4) = 1/(1+ \phi_4/\phi_3)$, which depends solely on $\phi_3/\phi_4$. We define the set of paths $\mathcal{S}$ that start by leaving $x$  and reach $x^*$ via an interaction with $RL$ without returning to $x$. The entropy change in the environment $\Delta S_{\rm env}[z(t)] $ due to each of these individual trajectories $z(t)$ is related to the probability of observing the forward pathway, and its reverse $\tilde z(t)$:\cite{Crooks1999, Seifert2011, Seifert2012}
\begin{equation}
\Delta S_{\rm env}[z(t)] = k \ln \left( \frac{p[z(t)|z(0)=x]}{p[\tilde z(t)|\tilde{z}(0)=x^*]} \right) - \Delta S_{\rm int}.
\end{equation}
\teoeditnewb{Here  $\Delta S_{\rm int}$ is an intrinsic entropy difference between macrostates $x$ and $x^*$ that arises because discrete biochemical states contain multiple microstates. It is a property of the states $x$ and $x^*$, rather than the transitions, and modifying it will not reduce overall dissipation in any steady-state network since the antagonistic transitions via $R$ must have an equal and opposite  $\Delta S_{\rm int}$.} We can then divide $\mathcal{S}$ into subsets $\mathcal{S}_i$ according to $\Delta S_{\rm env}[z(t)]$; all paths within ${\mathcal S}_i$ \teoeditnewb{generate the same entropy in the environment, $\Delta S^i_{\rm env}$}. Conceptually, different subsets might correspond to pathways that consume
different amounts of ATP. \teoeditnewb{A simple one-step pathway, as considered hitherto, is a special case of a system with a single $\mathcal{S}_i$.}

The probability of observing any pathway within $\mathcal{S}_i$ given an initial state $z=x$ is
\begin{equation}
P(\mathcal{S}_i|x) = \sum_{z(t)  \in \mathcal{S}_i} p(z(t)|z(0) = x).
\end{equation} 
Similarly,
$P(\mathcal{S}_{-i}|x^*) = \sum_{z(t)  \in \mathcal{S}_i} p(\tilde{z}(t)|\tilde{z}(0) = x^*)$ 
is the probability of observing a reverse of a pathway in ${\mathcal S}_i$ given an initial state $x^*$. By design, $\Delta S^i_{\rm env} + \Delta S_{\rm int}= k \ln(P(\mathcal{S}_i|x)/P(\mathcal{S}_{-i}|x^*))$. Further, the ratio of forwards to backwards fluxes is given by 
\begin{equation}
\frac{\phi_3}{\phi_4} = \frac{{\sum_i} P(\mathcal{S}_i|x) }{\sum_i P(\mathcal{S}_{-i}|x^*) }.
\label{accuracy1}
\end{equation}

An immediate consequence is that in systems with only one subset
of $\mathcal{S}$  in which all transitions have the same
$\Delta S_{\rm env} \equiv
  \Delta S^{\rm tight}_{\rm env} $, the ratio of forwards to backwards fluxes is fixed by
\begin{equation}
k \ln \left(\frac{\phi_3}{\phi_4} \right) = \Delta S^{\rm tight}_{\rm env}+ \Delta S_{\rm int}.
\label{tightly_coupled}
\end{equation}
Such a process is ``tightly coupled''. In our framework, being tightly coupled implies that each transition $x$ to $x^*$ is  associated with the breakdown of the same number of ATP molecules. In this case, regardless of the number of intermediate steps or possible pathways in a tightly-coupled process between two given states $x$ and $x^*$, 
the copy accuracy is unambiguously determined by the entropy change in the environment.
 Thus all tightly-coupled process with the same accuracy are associated with the same dissipation. Since the one-step process we have considered hitherto is tightly-coupled, all tightly-coupled processes have the same accuracy/efficiency trade-off as we have derived.

It is also possible to consider processes that are not tightly-coupled, with multiple subsets ${\mathcal S}_i$. In this case, the average entropy generated in the environment per $x \rightarrow x^*$ transition is given by 
\begin{equation}
\Delta S^{\rm multi}_{\rm env}= k \sum_i \frac{P^{\rm m}(\mathcal{S}_i|x) }{\sum_j P^{\rm m}(\mathcal{S}_j|x) } \ln \left( \frac{P^{\rm m}(\mathcal{S}_i|x) }{P^{\rm m}(\mathcal{S}_{-i}|x^*) } \right) -\Delta S_{\rm int},
\label{dS_multi}
\end{equation}
where we used the ``m'' superscript to denote probabilities for this specific multi-subset system. The accuracy of this process is still determined by Eq.\,\ref{accuracy1}:
$ \phi_3^{\rm multi} / \phi_4^{\rm multi} =\sum_i {P^{\rm m}(S_i|x)}/{P^{\rm m}(S_{-i}|x^*)}$. We can  compare this multi-subset process to a tightly-coupled process with the same accuracy: $\phi_3^{\rm tight} / \phi_4^{\rm tight} = \phi_3^{\rm multi} / \phi_4^{\rm multi}$. The key point is that since entropy generation and accuracy are unambiguously related for a tightly-coupled process (Eq.\,\ref{tightly_coupled}),  any tightly-coupled process between $x$ and $x^*$ with the same accuracy as the multi-subset generates 
\begin{equation}
\Delta S^{\rm tight}_{\rm env} + \Delta S_{\rm int} =
\\= k \ln \frac{\phi_3^{\rm multi}}{\phi_4^{\rm multi}} =  k \ln  \frac{{\sum_i} P^{\rm m}(\mathcal{S}_i|x) }{\sum_i P^{\rm m}(\mathcal{S}_{-i}|x^*) }.
\label{dS_tight}
\end{equation}
Combining Eqns.\,\ref{dS_multi} and \ref{dS_tight} thus yields
\begin{equation}
\Delta S^{\rm multi}_{\rm env} - \Delta S^{\rm tight}_{\rm env} = \sum_i P^{\rm m}_N(\mathcal{S}_i|x)  \ln \left( \frac{P^{\rm m}_N(\mathcal{S}_i|x) }{P^{\rm m}_N(\mathcal{S}_{-i}|x^*)} \right),
\label{KL}
\end{equation}
where the subscript \teoeditnewb{$N$ indicates normalization:} $P^{\rm m}_N(\mathcal{S}_i|x) = P^{\rm m}(\mathcal{S}_i|x)/\sum_j P^{\rm m}(\mathcal{S}_j|x)$, and $P^{\rm m}_N(\mathcal{S}_{-i}|x^*)$ is defined equivalently. Eq.\,\ref{KL} is a Kullback-Leibler divergence between the families of paths taken by forwards and backwards transitions in the multi-subset process, and is therefore necessarily non-negative. Thus 
\begin{equation}
\Delta S^{\rm multi}_{\rm env} - \Delta S^{\rm tight}_{\rm env} \geq 0.
\end{equation}
 Similarly, paths that begin and end in $x$ or $x^*$ generate no entropy for a tightly coupled process, whereas the entropy generation of these paths is non-negative in general. For a system operating in steady state, increased entropy deposited into the environment implies less efficiency. Therefore no process of a given copy accuracy is more efficient than a tightly-coupled one. The limits derived for a one-step process, \teoeditnewb{a special case of a tightly-coupled process}, are therefore general. 

Our derivation is related to the proof that the estimate of
dissipation obtained from the irreversibility of a coarse-grained
trajectory gives a lower bound on the true entropy generation
\cite{Kawai2007,Gomez-Marin2008}.  The results are fundamentally distinct, however. We could consider coarse-graining a complex copy process so that all states were now $x$ or $x^*$. However, the result would not be a simple one-step process with the same accuracy and lower dissipation; the dynamics would be non-Markovian and reflect the underlying complex process. To state that a true one-step process could reproduce the accuracy at the same cost as  estimated from the coarse-grained description would be to assume the conjecture that is to be proven.


\subsection{Biochemical implementation of an optimal device and protocol.}
We have argued that no   biochemical copying network, operating autonomously and directly powered by a non-equilibrium fuel supply,
can reach the thermodynamic bound on efficiency for general input data. We now consider whether this is a fundamental 
property of biochemical reactions, or whether biomolecules could in principle act as thermodynamically optimal bits.

There are two principal differences between cellular biochemical networks and optimal protocols
\cite{Bennett1982,Bennett2003,feynman1998feynman}. Firstly, cellular
networks operate  continuously, rather than taking a series of discrete
measurements with external clocking. Secondly,  as emphasized above, they involve  no manipulation of \teoresp{(free)-}energy levels \teoresp{over time}, as illustrated in Fig.\,\ref{fig:landscapes}. 

\teoeditnewb{Concerning the first difference, in Supplementary Discussion 5 we show
that a clocked analogue of the cellular push-pull motif (illustrated in Supplementary Fig. 4) gives a work per copy identical to the continuous case. 
 This is  because, despite being operated in a clocked fashion, the device is still out of equilibrium and functions at constant chemical potential of fuel molecules.  The fact that cellular networks operate in a stochastic continuous manner  rather than a clock-like fashion is not the fundamental reason why they cannot reach the bound on the energy cost of a copy operation.}

Functioning out-of-equilibrium is necessary for a device operating at constant chemical potential of fuel; if the reactions were in equilibrium, the receptor (which is a catalyst) could not influence the yield of $x/x^*$. We now show that a system  driven by quasistaic manipulation of ATP, ADP and P concentrations could reach the thermodynamic bound, \teoresp{confirming that the autonomous network is inefficient due to its static free-energy levels}. \teoedit{To operate in the quasistatic limit, $RL$ and $R$ must be long-lived; in practice, constitutively active kinases and phosphatases could be used.}

\begin{figure}
\centering
\includegraphics[width=8.5cm]{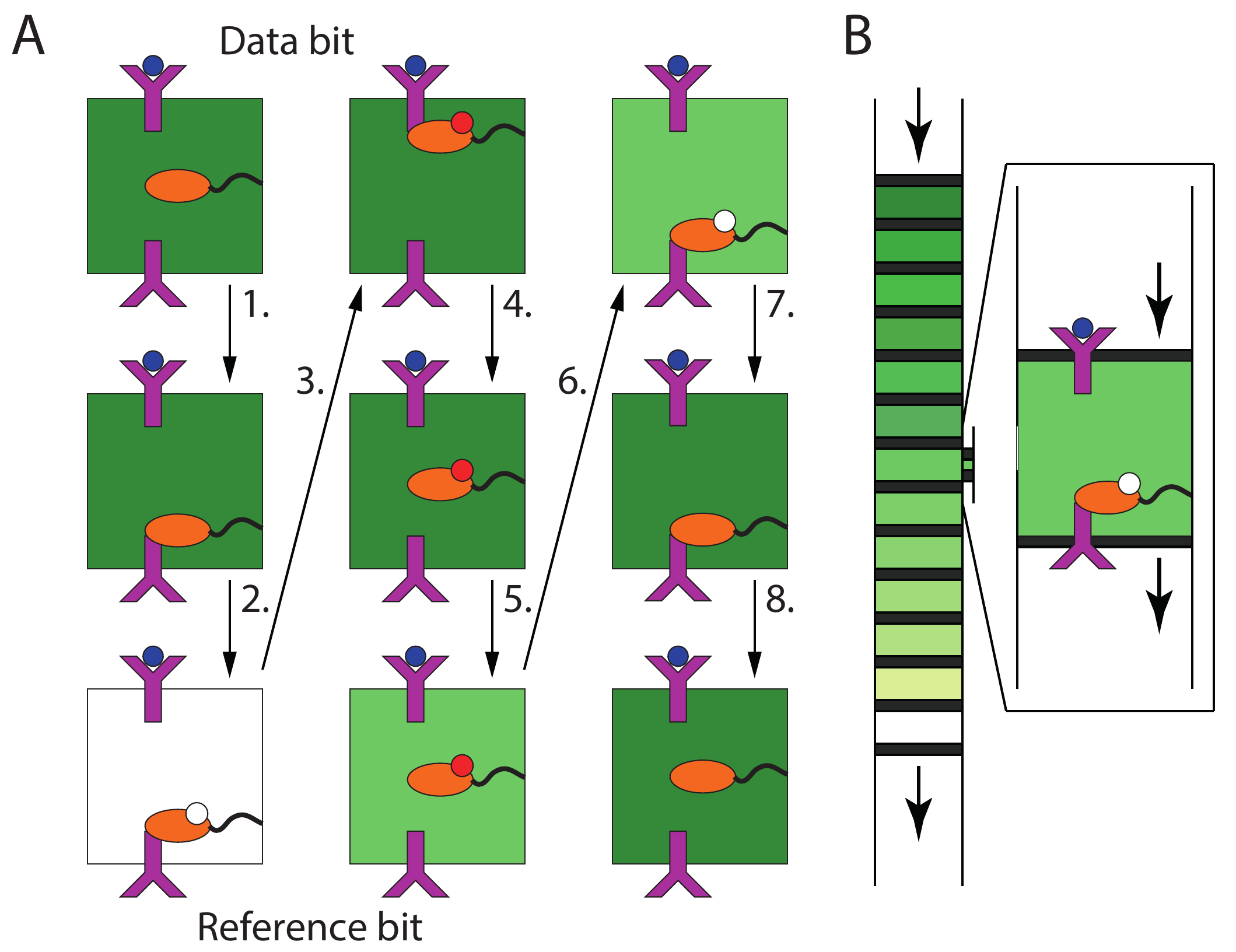}
\caption{A biochemical implementation of an optimal device and copy protocol. The cycle is illustrated in (A), and the steps are explained in the text. \teo{The system contains two receptors: one acting as a reference bit in state $R$ for resetting and a receptor acting as a data bit in state $RL$ or $R$. These receptors are attached to either end of a reaction volume. A readout molecule (memory bit) is tethered to the side of the reaction volume. The colour of the reaction volume indicates the chemical potential of fuel molecules in solution: it is dark when $\Delta G_p, \Delta G_d$ are large and negative and white when $\Delta G_p, \Delta G_d =0$ }. (B) A device for implementing this cycle. A small reaction volume is coupled to a piston containing a series of reservoirs of varying ATP, ADP and P content. Similarly, the receptors and readout can be brought in and out of proximity by manipulation of a second piston.  \label{fig:bio_quasi}  }
\end{figure}

We consider a device and measurement cycle as illustrated schematically in Fig. \ref{fig:bio_quasi}. The key ingredients are the possibility of manipulating the concentrations of ATP, ADP and P in the vicinity of the readout, and the ability to bring the readout into or out of close proximity with receptors. We consider the same receptor/readout reactions as in Eq. \ref{reactions}, and now define the free energy changes of reaction $\Delta G_p$ for the phosphorylation of $x$ by $RL$ and $\Delta G_d$ for the dephosphorylation of  $x^*$ by $R$. These quantities are given by
\begin{equation}
\begin{array}{l l l}
\Delta G_p & = & \mu_{\rm ADP} - \mu_{\rm ATP} + \Delta G_{x/x^*},\\
\Delta G_d & = & \mu_{\rm P} - \Delta G_{x/x^*},\\
\end{array}
\label{eq:DGd and DGp}
\end{equation}
in which $\Delta G_{x/x^*}$ quantifies the intrinsic stability of $x$ and $x^*$ (assumed to be independent of the chemical potentials of ATP, ADP and P \cite{Seifert2011}). The reactions can thus be manipulated by controlling  $\mu_{\rm ATP}, \mu_{\rm ADP}$ and $\mu_{\rm P}$, for example by coupling of the system to a series of reservoirs (Fig. \ref{fig:bio_quasi}\,(B)). We outline an optimal cycle below, and calculate the chemical work done on the readout subsystem by the reservoirs (the average number of reactions of a given type, multiplied by the associated chemical work, integrated over the whole process) in Supplementary Discussion 6. Throughout, we assume that receptor states are stable, and that reactions cannot occur without a catalyst. The eight steps of the cycle are intended to be closely analogous to typical computational protocols;\cite{Bennett1982,Bennett2003,feynman1998feynman} for subtleties involved in this comparison, see  Supplementary Discussion 6. The system contains a readout (the memory bit) a receptor of known state $R$ (a reference bit) and a receptor in either $R$ or $RL$ (the data bit).

The readout begins coupled to a buffer with $\Delta G_p, \Delta G_d= -\Delta G_r$ ($\Delta G_r$ assumed to be large and positive). There is no receptor in close proximity, but the readout has been reset (equilibrated \teoqq{at the end of a previous measurement cycle} by a receptor in the $R$ state), and therefore is in state $x$ with probability $1/(1+ \exp(-\Delta G_r/kT))$. In fact, the accuracy of this reset does not influence the cost of the cycle (see Supplementary Discussion 6). 
\begin{enumerate}
\item The readout is brought into close proximity with a receptor of known state $R$. 
\item  $\Delta G_p, \Delta G_d$ are slowly (quasistatically) raised from $-\Delta G_r$ to 0. Steps 1 and 2 allow the state of the memory to be reversibly uncorrelated from the reference receptor, prior to the measurement.
\item The readout is brought into close proximity with a receptor of unknown state (ligand-bound with probability $p^\prime$). $\Delta G_p, \Delta G_d$ are then slowly lowered to $-\Delta G_s$. In this step, the state of the readout is set to match that of the receptor with accuracy $s = 1/(1+\exp(-\Delta G_s/kT))$.
\item The readout is moved away from the receptor. 
\item $\Delta G_p, \Delta G_d$ are set to $\Delta G_{\rm off}$. This stage constitutes the end of the copy subprocess; if the unknown receptor is in state $RL$, then the readout is in $x^*$ with probability $s = 1/(1+\exp(-\Delta G_s/kT))$. Similarly, if the unknown receptor is in state $R$, the readout is in $x$ with probability $s = 1/(1+\exp(-\Delta G_s/kT))$.
\item We now decorrelate the memory and data bits at a non-zero bias between the two readout states, $\Delta G_p,\, \Delta G_d = \Delta G_{\rm off}$. To reach the fundamental thermodynamic bound, $\Delta G_{\rm off}$ must be chosen carefully, and depends on $p^\prime$ and $s$. To decorrelate,
the readout is brought into close proximity with a known receptor of state $R$. The readout relaxes to a state reflective of  $\Delta G_{\rm off}$ via the reaction $ R + x^*  {\rightleftharpoons} R + x + {\rm P}$. 
\item The readout molecule is reset by quasistatically lowering $\Delta G_p, \Delta G_d$ to $-\Delta G_r$ from $\Delta G_{\rm off}$, returning it to a state dominated by $x$.
\item The readout is separated from the reference bit, returning the system to the initial state.
\end{enumerate}
The readout and receptor are in the same state at the start and finish of the cycle; thus the net chemical work of the reservoirs equates to the free energy dissipated by the entire system. As shown in Supplementary Discussion 6, minimizing dissipation with respect to $\Delta G_{\rm off}$ at fixed $p^\prime$ and accuracy $s$ (fixed by $\Delta G_s$) gives a chemical work per copy equal to the mutual information generated by a measurement.
In the special case $p^\prime=1/2,\,\Delta G_s \rightarrow \infty$, $
{w_{\rm chem}} \rightarrow kT \ln 2$, as expected. \teoeditnewb{A single readout could also be made to copy multiple receptors sequentially -- using, for example, a series of receptors anchored to a polymer, as in Supplementary Fig. 4}. We note that it is also possible to construct an optimal cycle if the receptor is only catalytically active in the $RL$ state -- see Supplementary Fig. 5 and Supplementary Discussion 7.

\section{Discussion}
We have described a canonical cellular readout network rigorously in terms of computational copy
operations, \teoeditnewb{thus demonstrating that the system is indeed bound by the 
thermodynamics of computation.} \teoeditnew{For a general distribution of input data, the cellular network
cannot reach this fundamental
  limit of efficiency}. Unlike optimal
  computational protocols, the thermodynamic driving force used to
  push the memory device between its states is not introduced
  quasistatically. Instead, a continuously operating autonomous
  network must have a constant thermodynamic  \teoresp{discrimination between correct and incorrect copy outcomes over time}.
 Even in the limit where the driving force and
  hence the accuracy become vanishingly small, the cellular system
is not thermodynamically optimal. In the regime of high
  accuracy, the difference is larger: while optimal protocols can reach 100\%
  accuracy for a finite cost of $kT \ln(2)$, cellular networks can
  only achieve 100\% accuracy for a cost that diverges. Nonetheless, achieving 99\% 
accuracy \teoeditnew{for an unbiased distribution of input data requires less than $4 kT
  \ln 2$ of dissipation per copy cycle, and the relative performance of the biochemical network
 for biased input data can be even better. Although $kT$ sets an energy scale,
it is not obvious {\it a priori} that the numerical factors 
should be so low. For example, the recently-derived 
``thermodynamic uncertainty relation'' \cite{Barato2015} --  in which the cost of achieving a relative uncertainty $\epsilon$
in the number of steps of a biomolecular process was shown to be at least $2kT/\epsilon^2$ -- gives $20000 kT$ for 99\% accuracy. }

Not only can this canonical cellular signalling system get remarkably close to the 
fundamental bound for efficiency of copying at relatively high accuracy,  it can do so at an arbitrarily high absolute copy rate. Further, the system is autonomous, and so there is no need to consider the intrinsic costs of applying a time-varying
yet stable control to a bit, as must be done in typical protocols \cite{Gopalkrishnan2015,Machta2015}. The canonical biochemical network also naturally adapts to
high or low levels of ligand-bound receptors, reducing its dissipation
per copy \teoedit{cycle} in a way that standard quasistatic protocols
cannot achieve without feedback. \teoresp{The remarkable possibilities of this biochemical network not only show that the thermodynamic limits of computation are genuinely relevant to practical systems, but emphasise that biological systems are an excellent environment to rigorously investigate these limits in a concrete, autonomous setting.}

\teoresp{Our mapping emphasises the cause of the minimal thermodynamic
  dissipation for the readout network. Dissipation does not occur
  because the memory is ``erased'' \cite{Mehta2012}---erasure itself
  is not intrinsically irreversible, and no distinct erasure step is
  present. Rather, the stable correlations generated between
  non-interacting systems by copying are not used to extract
  work \cite{Fahn1996}. Other biochemical processes, such as
  ubiquitination, transcription, translation and replication, result
  in correlations between degrees of freedom that are not maintained
  by direct interactions.  Indeed, although we were motivated by the
  time integration of receptors by readouts, our analysis directly
  applies to other push-pull networks. Our work suggests that the
  thermodynamics of these persistent correlations is a central
  paradigm through which to understand this class of systems.}

Our analysis has revealed that biomolecules can in principle be used to implement protocols that achieve
the thermodynamic bound, and we have
provided an example. The key difference from the canonical
cellular network is the manipulation of concentrations of ATP,
ADP and P \teoresp{over time, during the course of the measurement}. If these manipulations are performed slowly enough,
reactions (except decorrelation of readout and receptor) can be
performed reversibly as reactants/products are gradually
stabilised/destabilised with respect to each other. 
We thus
propose a new class of systems in which the fundamental thermodynamics
of computation can be explored, to complement experiments done with
optical or \teoedit{electrostatic feedback} traps
\cite{berut2012experimental, Jun2014} and magnetic systems \cite{Hong2014}. \teoedit{Our approach involves
  manipulating biomolecules by adjusting the chemical potential of
  fuel molecules}. Our experimental system is particularly promising because the
dissipation could in principle be measured directly for a large number
of devices acting in parallel rather than inferred from positional
trajectories as it is done for the optical \teoedit{or feedback} traps. 
A second advantage of the proposed setup is that the store of free energy used to perform work on
the system -- the chemical fuel -- is explicit. It is clear exactly how free energy is transferred to and absorbed back from the
bit under study. In other analyses, the explicit mechanism by which work is transferred between a bit and a store of free 
energy is implicit. It is usually assumed that the store of free energy can supply and absorb work efficiently, 
even if the operations on the bit itself are irreversible. In practice, however, work is typically supplied in a highly irreversible fashion such as via 
lasers \cite{berut2012experimental, Jun2014},
 and any work done by the bit is lost rather than stored.

\teoqq{In an experimental realization, it would be natural to treat the reservoirs and memory together as an extended system thermally coupled to the outside world.  In this case the chemical free energy dissipated during measurement is not equal to the heat exchanged between the extended system and the outside world \cite{Seifert2011} -- the increased entropy of the universe is instead manifest in a less uneven distribution of ATP, ADP and P between reservoirs. It would therefore be most natural to measure dissipation through the changing concentrations of ATP, ADP and P as the reservoirs exchange molecules \teoedit{-- perhaps through radioative labelling of phosphates}.} Further, it should be possible to perform full measurement cycles and probe the link between information loss and irreversibility. By measuring the state of the readout using, e.g., FRET, it would also possible to test the generalized Jarzinksy equality, which shows that the state of the system can be changed more efficiently by exploiting knowledge of the state of the system \cite{toyabe2010experimental,Koski2014,Roldan2014}.

\teo{ The fact that cells employ thermodynamically inefficient
  out-of-equilibrium circuits}, \teoresp{despite energy budgets being an important consideration 
in evolutionary fitness,}
 highlights the constraints
under which they function. Cells do not have infinite time to
perform a measurement,
meaning that quasistatic manipulations are infeasible. Further, our
quasistatic protocol requires that a
readout molecule exclusively encounters either ligand-bound {\it or}
ligand-free receptors during each copy process. This requires that the
ligand and ligand-free states of the receptors are stable on the time
scale of the measurement cycle, and also that receptors in both states are not accessible to a
single readout. In reality, the finite lifetime of receptor states
places limits on the measurement time, and cells typically have
multiple receptor molecules with which any one readout molecule can interact.  The
quasistatic cycle also necessitates coordinating the separation between readouts and receptors; although this is
not inconceivable within a cell, it would require elaborate
machinery. Perhaps most importantly, however, the quasistatic protocol
requires manipulation of the concentration of chemical fuels; this manipulation must change chemical potentials by
several $kT$ to be effective. Given the relatively small number of
chemical fuels available, and their extensive use in a
range of systems, it would be very surprising if the cell manipulated
chemical potentials purely for the sake of measurement efficiency. 

The final observation may also explain why cells use such a strong chemical driving 
(the  hydrolysis of ATP typically provides approximately $20kT$, deep into the low error regime),
rather than more efficiently taking measurements of only slightly lower accuracy \cite{govern2014}.
Indeed, the most efficient strategy from the perspective of sampling is to make many low-accuracy copies \cite{govern2014}; 
this, however, requires time (for the measurements to be independent) and readout molecules (to store the measurements), resources
which are not free for the cell. 
The design of more efficient copying architectures may be relevant in
 synthetic biology and biological engineering in
which the constraints and goals are distinct from those of natural
systems.

In this work we have focussed on a single-step copy process. However, we have also argued that more complex copy processes, including multiple bound receptor/readout states or even multiple pathways in which different amounts of ATP can be consumed, cannot be equally accurate at a lower cost. Indeed, we have shown that all ``tightly-coupled'' processes in which all transitions consume the same amount of chemical fuel are equally efficient, and all others cannot perform better. At a physical level, it is intuitive that complex process can be less efficient---they naturally allow for dissipative cycles. Similarly, longer pathways are not helpful because one cannot make an irreversible process less irreversible by breaking it into many small steps whilst keeping the overall driving force fixed.
For processes such as kinetic proofreading, however, in which chemical
fuel drives reactions out of equilibrium, complex or longer reaction
pathways can allow the same outcome at a lower cost \cite{Qian2006, Lang2014, Ehrenberg1980,Murugan2012}. The
crucial difference is that in a copy process, the metric of accuracy
is exactly the ratio of forwards to backwards transition
probabilities, which is the quantity directly influenced by fuel
consumption. More fuel consumption always improves the metric. For
other tasks, fuel consumption is necessary, but other factors can also
influence performance. In kinetic proofreading, the relevant metric is
the relative occupancy of a binding site by two different ligands
\cite{Qian2006,Ehrenberg1980}. This quantity is influenced by the
strength of chemical driving, but it is also limited by the number of
intermediate states at which there is an opportunity to discriminate
between ligands. Thus it can be beneficial to consider multi-stage
proofreading \cite{Murugan2014}.

\teoedit{Our analysis of the cellular network used the
  mean-field limit, which becomes accurate when the receptor
  correlation time $\tau_c$ is shorter than the relaxation time
  $\tau_r$ of the readout network. Interestingly, this is precisely
  the optimal regime for sensing  \cite{govern2014}, because  the system can take multiple $\tau_r / \tau_c > 1$
  concentration measurements per receptor molecule.  In this
regime, the readout molecules do not track the fluctuations
in the receptor state, but, collectively, average it. As a
result, the ``learning rate'' \cite{Barato2014} between the readout
and the receptor is actually zero in this limit (see
Supplementary Discussion 1). While the opposite regime $\tau_r / \tau_c < 1$ is
  detrimental for the mechanism of time integration, we do note that
  the work per measurement is less.  This is because in this regime
  the measurements become correlated, and taking correlated
  measurements requires less work for a given desired accuracy.
 A full analysis of this regime is the subject of further work. \teoresp{Similarly, we have not considered} the consequences
 of spontaneous reactions not mediated by kinases and phosphatases. \teoresp{In the context of copying, these reactions equate to spontaneous thermalisation of bits, which could be incorporated into our mapping.} }

\emph{Acknowledgements}: We thank Giulia Malaguti for a critical reading
  of the manuscript and Andrew Turberfield for helpful discussions. 
 This work is part of the research programme of the
  Foundation for Fundamental Research on Matter (FOM), which is part
  of the Netherlands Organisation for Scientific Research
  (NWO). TO is supported by a Royal Society University Research Fellowship.

\bibliographystyle{unsrt} 
\bibliography{sensingbib2}

\begin{appendix}

\begin{widetext}

\pagebreak[4]

\section{Supplementary Figure 1: computational copy cycle }
\begin{figure}[h!]
\includegraphics[width=8.5cm]{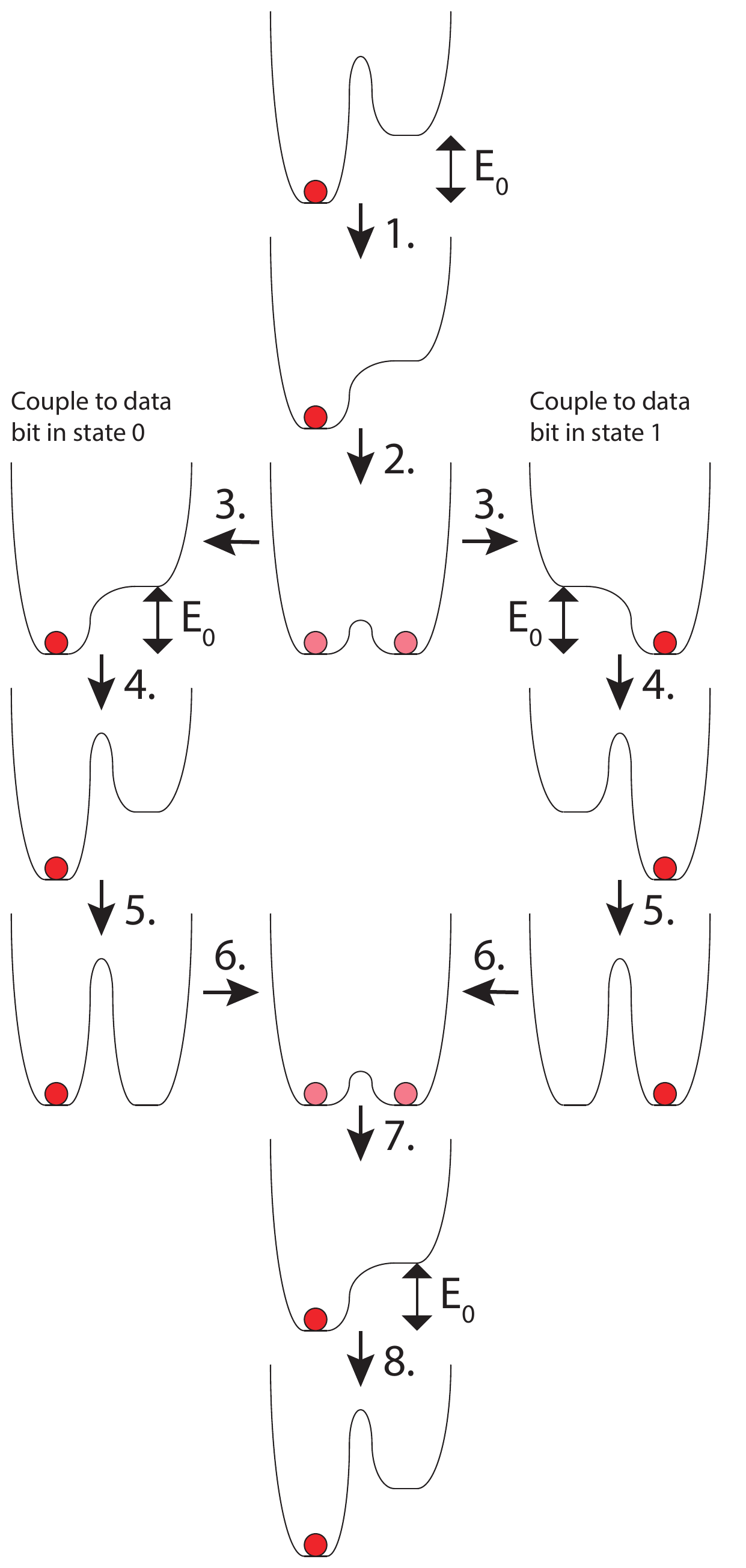}
\caption{A canonical computational measurement/copy cycle with a particle in a double well potential representing a memory bit. The particle starts in state 0 (the left well) with the energy of the right well raised by $E_0 \gg kT$ due to coupling of the memory to a reference bit. Step 1: the barrier between wells is lowered. Step 2: The memory is slowly decoupled from the reference bit, allowing the right well to fall. The particle remains in equilibrium between the two wells. Step 3: the memory is gradually coupled to a data bit of unknown state, raising either the left or the right well. Step 4: the barrier between the wells is raised. Step 5: the memory and data bits are decoupled, completing the copy stage of the protocol. Steps 6-8 are performed prior to the next copy to return the memory bit to state 0. Step 6: the barrier between wells is lowered, allowing the particle to equilibrate between the wells. Step 7: the memory is gradually coupled to the reference bit, and the particle is returned to the left well. Step 8: the barrier is raised, returning the memory bit to its initial state 0. Steps 1-5 constitute the copy protocol, step 6 is decorrelation, and steps 7 and 8 are the reseting of the memory. 
\label{canonical protocol}}
\end{figure}

\pagebreak[4]

\section{Supplementary Figure 2: efficiency for bits with unequal sampling probabilities}
\begin{figure}[h!]
\includegraphics[width=8cm, angle=-90]{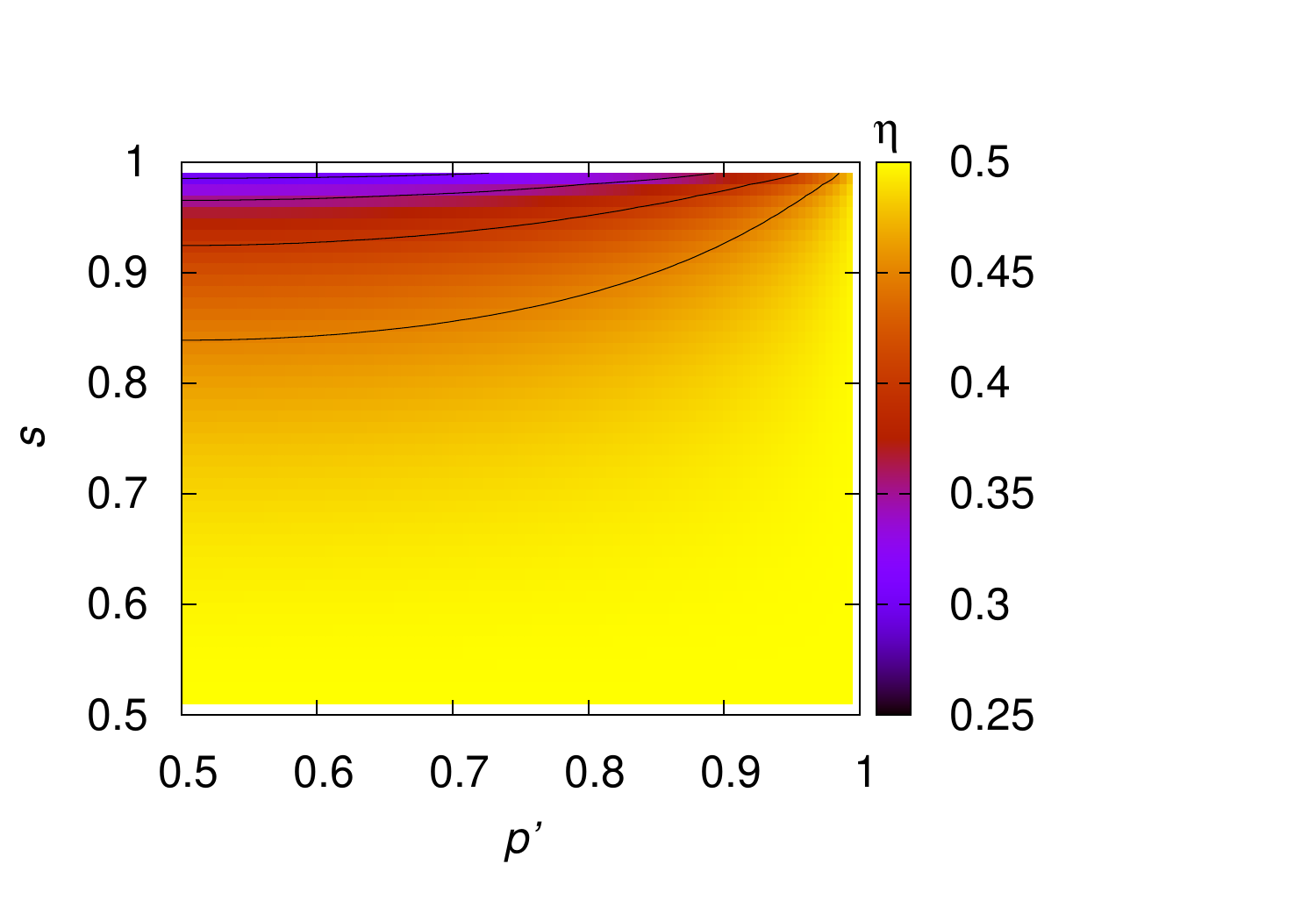}
\caption{ Efficiency $\eta$ as a function of $p^\prime$ and $s=s_R=s_{RL}$. Note that $\eta$ does not exceed 0.5, and that moving $p^\prime$ away from 0.5 always increases $\eta$  at fixed $s$. A mirror image is obtained for $p^\prime<0.5$.
\label{eta_p_s}}
\end{figure}

\pagebreak[4]

\section{Supplementary Figure 3: efficiency for bits of unequal copy accuracies}
\begin{figure}[h!]
\includegraphics[width=6cm,angle=-90]{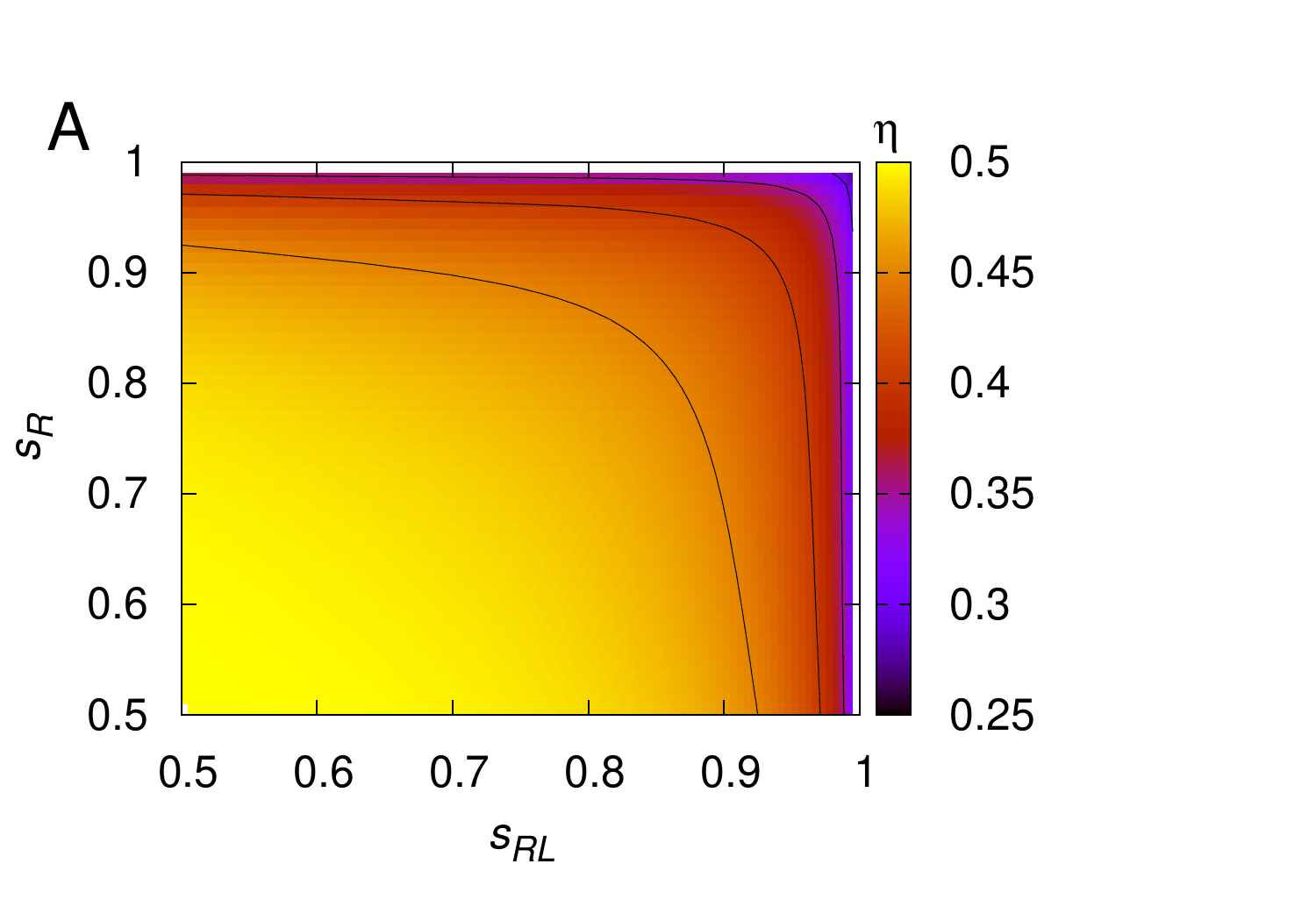}
\includegraphics[width=6cm,angle=-90]{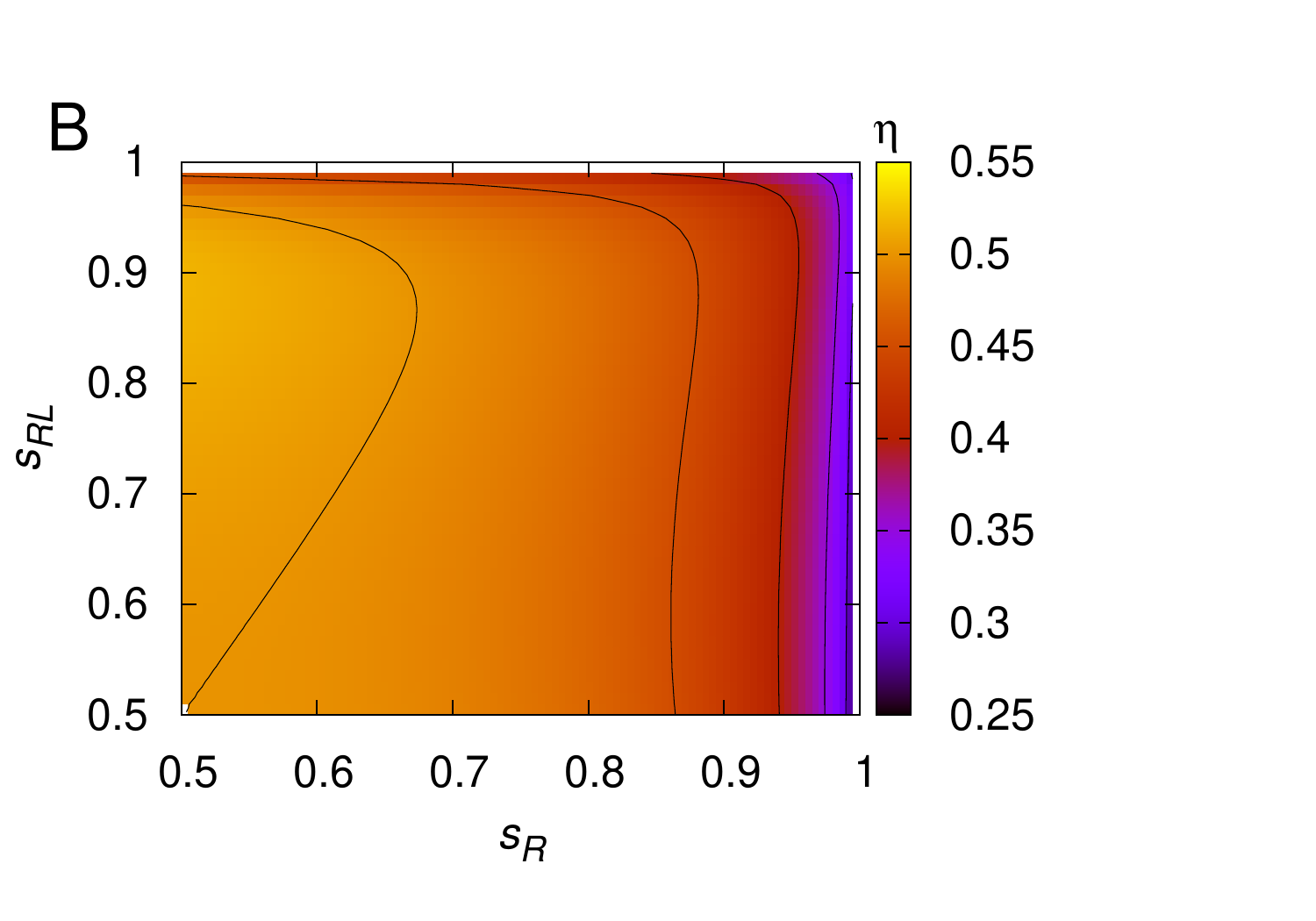}
\includegraphics[width=6cm,angle=-90]{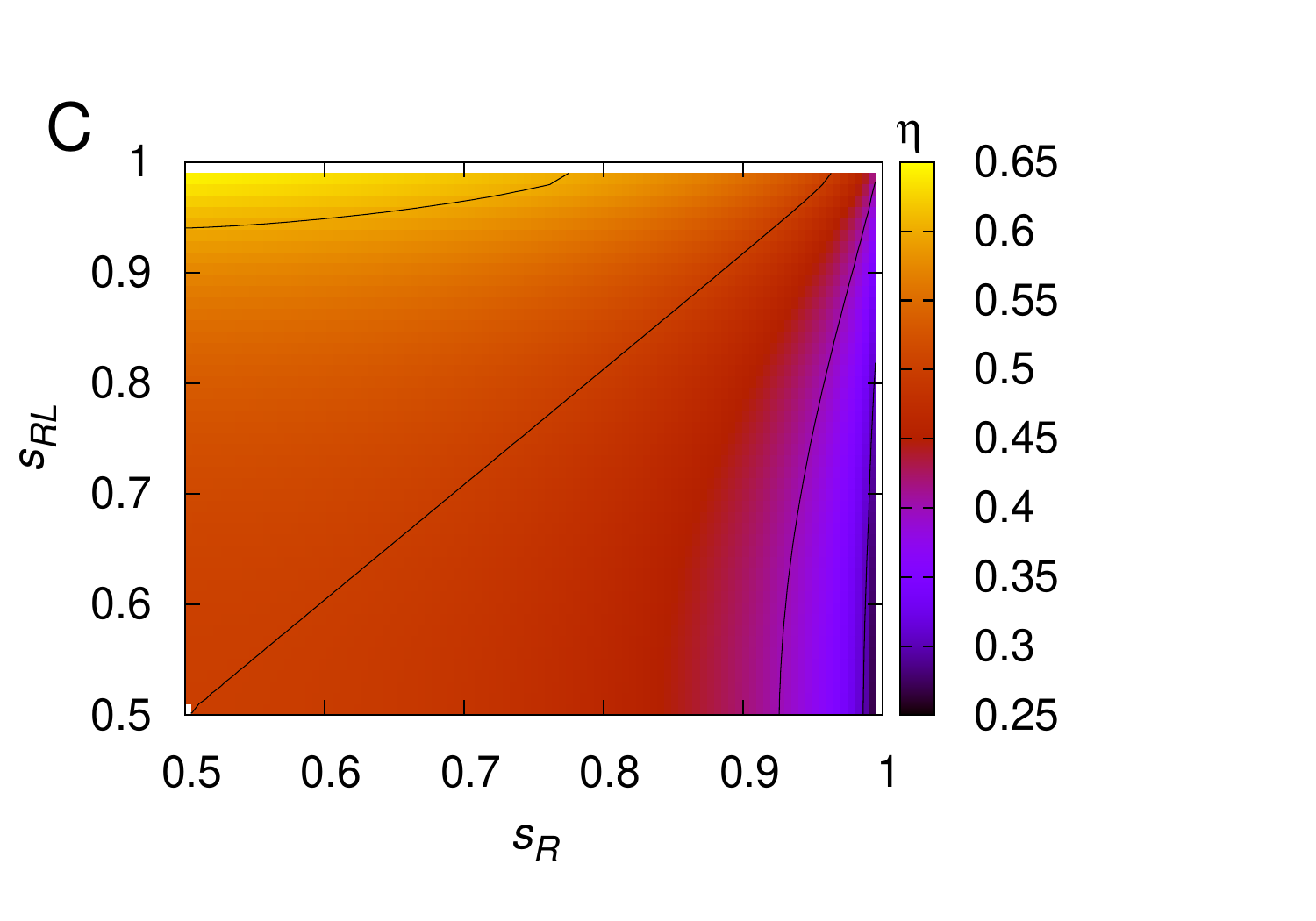}\\
\includegraphics[width=6cm,angle=-90]{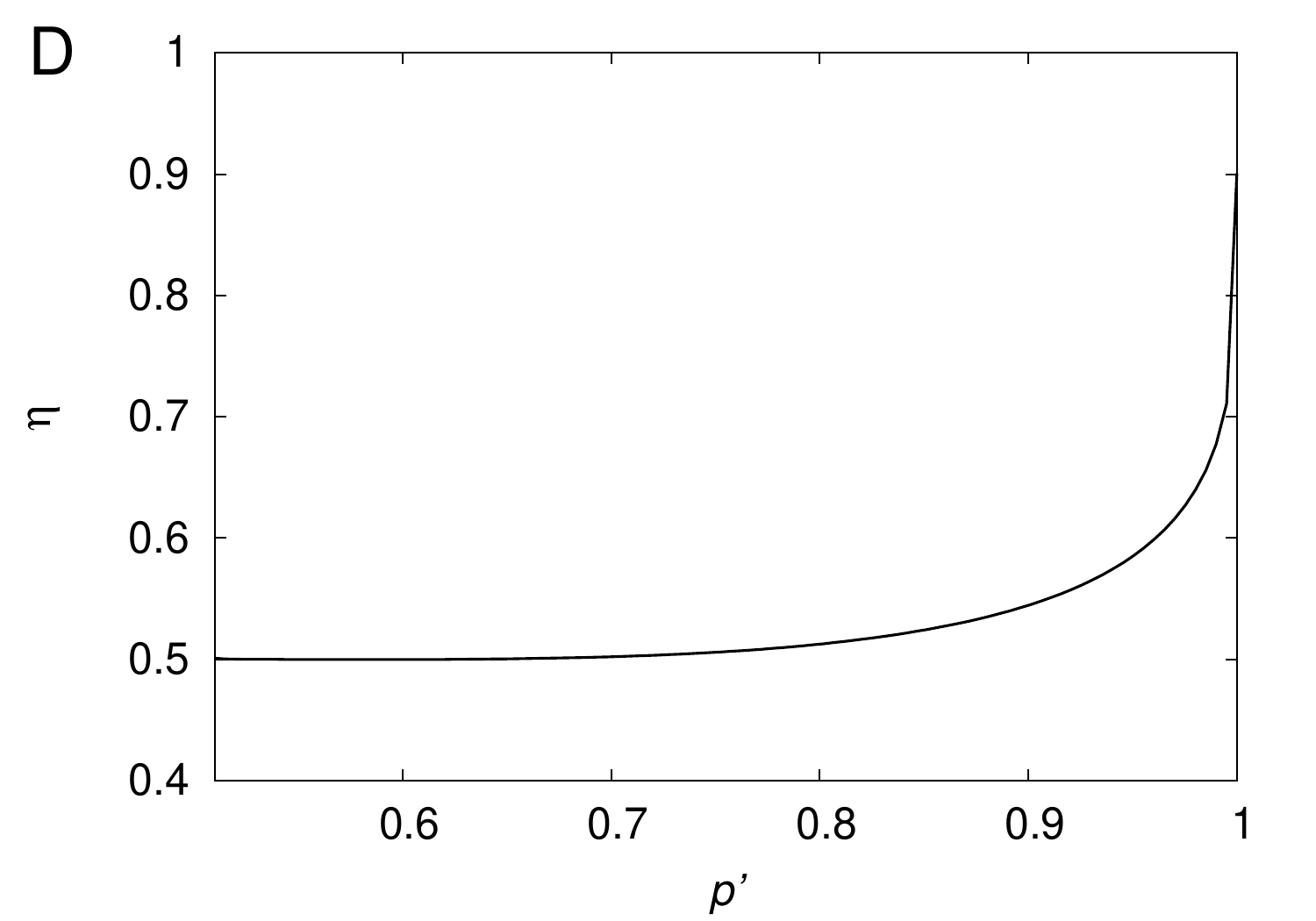}
\caption{ Efficiency of the biochemical network $\eta$ as a function of $s_R$and $s_{RL}$ for (A) $p^\prime=0.5$, (B) $p^\prime=0.8$ and (C) $p^\prime=0.98$. Note the substantial region in which $\eta$ exceeds 0.5 that appears as $p^\prime$ grows, centred around $s_{RL} \approx p^\prime$ and $s_R \approx 0.5$. Also note that as  $s_{RL} \rightarrow 1$ at fixed $p^\prime$, $\eta$ is small. (D) Plot of $\eta$  against $ p^\prime=s_{RL}$ at $s_R=0.5$. Note that convergence on unity only occurs at extreme values of $s_{RL}= p^\prime$. \label{eta_sR_sRL}}
\end{figure}

\pagebreak[4]

\section{Supplementary Figure 4: a clocked biochemical device and protocol}
\begin{figure}[h!]
\includegraphics[width=7cm]{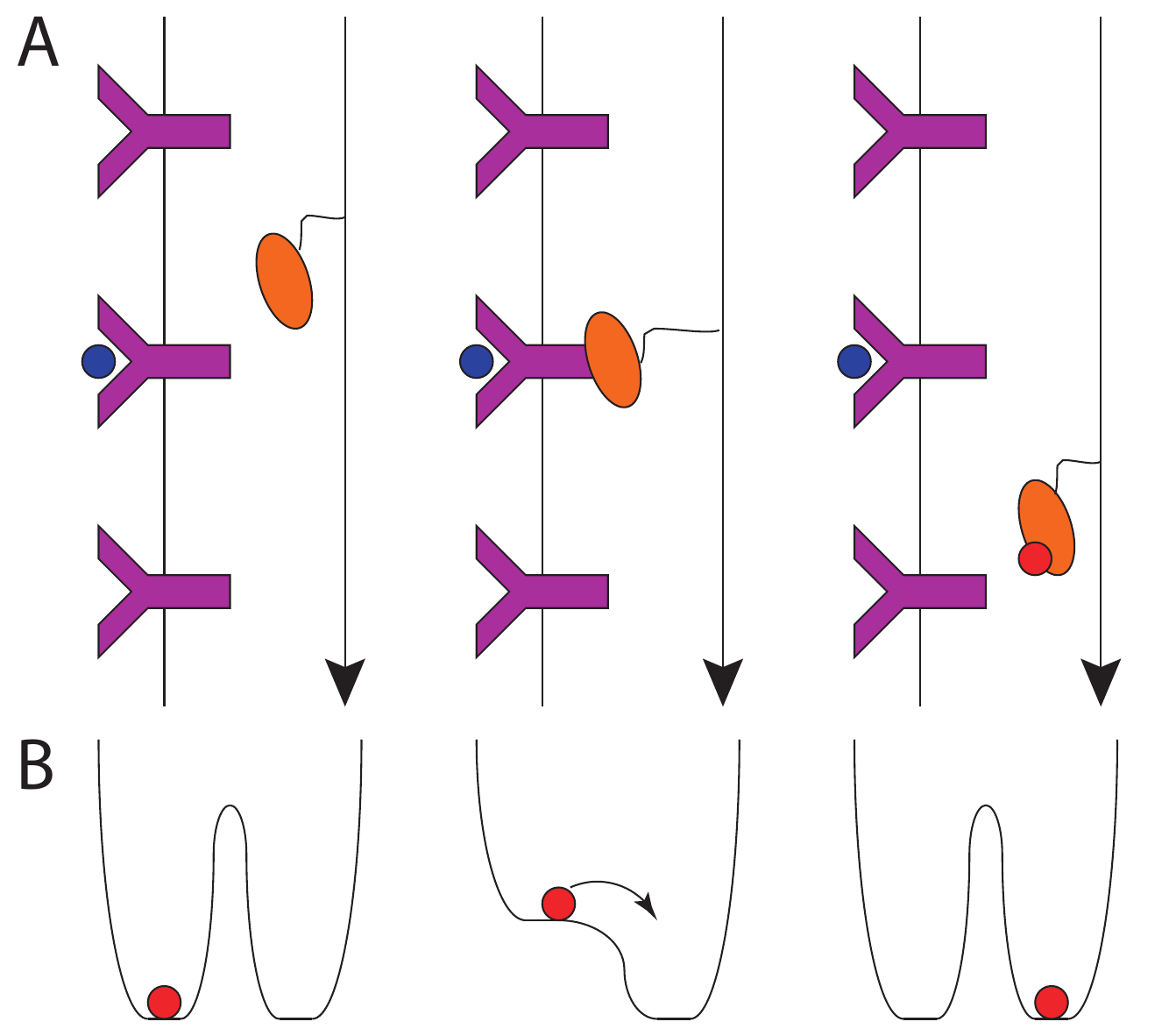}
\caption{An artificial biochemical system in which measurements
 are performed in a clocked fashion, analogous to computational
 models. (A) Schematic of the system -- the readout, attached to a
 polymer by a tether, is moved past a series of receptors. (i) The
 memory is between receptors, and cannot interact with them. (ii) The
 memory is brought into proximity with a receptor, allowing catalysis
 and copying. (iii) The memory is moved on, and cannot interact with
 any receptors. It retains the state of the previous receptor until
 the next measurement. (B) \teoedit{An interpretation of the process in
  (A) in a
two-state energy landscape picture. In this analogy, the left well
 corresponds to state $x$ and the right well to $x^*$. The presence of an enzyme 
simultaneously lowers the barrier between $x$ and $x^*$ and shifts the equilibrium.}
\label{biochem protocol 1}}
\end{figure}

\pagebreak[4]

\section{Supplementary Figure 5: An optimal biochemical copy cycle with a mono-functional receptor}
\begin{figure}[h!]
\centering
\includegraphics[width=8.5cm]{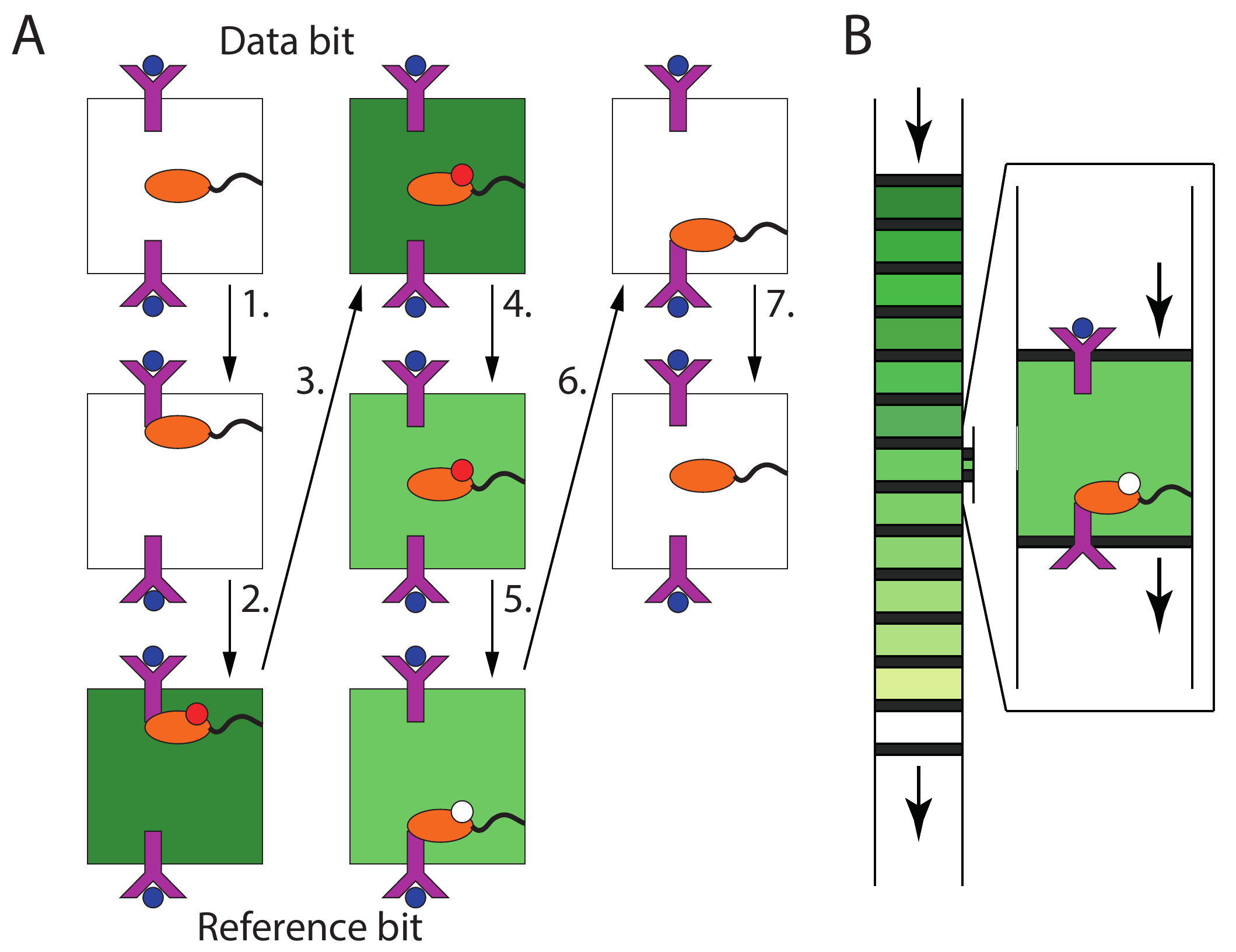}
\caption{A biochemical implementation of an optimal copy device and protocol, in which only the $RL$ state of the receptor is catalytically active. The cycle is illustrated in (A). \teo{As in the main text, the system consists of receptors acting as data and reference bits attached to the ends of a reaction volume, and a tethered readout (memory bit). The reference bit is a receptor in state $RL$, the data bit is a receptor in either $R$ or $RL$ (illustrated here as as $RL$).} Initially, the readout is not in close proximity to any receptor, and is in a solution in which $\Delta G_p =\Delta G_s$ is large and positive (the reaction in Eq. \ref{eq:just kinase} is driven to the left), indicated by the light color of the reaction volume. \teo{Its is unphosphorylated due to prior coupling to the reference bit.} Step 1: the readout is brought into close proximity with the receptor acting as a data bit. Step 2: the solution is quasistatically changed from $\Delta G_p  = -\Delta G_s$ to $\Delta G_p =\Delta G_s$  (changed conditions are shown by the darker color). \teo{If the receptor is in state $RL$, the readout tends to be phosphorylated.}  Step 3: the readout is separated from the receptor acting as a data bit. Step 4: the solution is quasistatically changed to $\Delta G_p = - \Delta G_{\rm off}$, the free energy of reaction at which decorrelation will take place. Step 5: the readout is brought into contact with the receptor acting as a reference bit to allow decorrelation with the data bit. Step 6: the readout is reset by quaistatically changing conditions to $\Delta G_p = \Delta G_s$. Step 7: the readout is separated from the reference bit, restoring the initial state. (B) A device for implementing this cycle. A small reaction volume is coupled to a series of reservoirs of varying ATP, ADP and P content. The current reservoir can be changed by pushing a piston. Similarly, the receptors and readout can be brought in an out of proximity by manipulation of a second piston. }
 \label{fig:bio_quasi2}  
\end{figure}

\teoedit{
\section{Supplementary Discussion 1: The master equation of the biochemical network}
The model presented in Eq. 2 of the main text defines a master equation for the variables $x^*$ and $RL$, the number of phosphorylated readouts and ligand-bound receptors, respectively.
\begin{equation}
\begin{array}{c}
\frac{{\rm d} P(x^*, RL)}{{\rm d} t} =  - ( [L]k_1 (R_T- RL) +k_2 RL) P(x^*, RL) \\
\\
+[L]k_1 (R_T- (RL-1)) P(x^*, RL-1)  +k_2(RL+1) P(x^*, RL+1)  \\
\\
- \left((k_3 \frac{RL}{V} + k_5  \frac{R_T - RL}{V})  (x_T-x^*)   + (k_4 \frac{RL}{V} + k_6  \frac{R_T - RL}{V})  x^*  \right)  P(x^*, RL) \\
\\
+( k_3 \frac{RL}{V} + k_5  \frac{R_T - RL}{V} )  (x_T-(x^*-1)) P(x^*-1,RL) + ( k_4 \frac{RL}{V} + k_6  \frac{R_T - RL}{V} )  (x^*+1) P(x^*+1,RL), 
\end{array}
\label{eq:master}
\end{equation}
in which $V$ is the system volume. We assume that the ligand concentration is large enough that $[L]$ is uninfluenced by $RL$ -- thus $\tilde{k_1}= k_1[L]$ is a constant. We also assume that bound states formed between
  receptors and downstream proteins are short-lived compared to the
  typical time between binding events, so that all receptor/readout
  reactions are essentially instantaneous. This is equivalent to
  assuming that the probability of finding the memory in between the
  wells (at the top of the barrier) is negligible in the
  computational device in Supplementary Figure 1. In this work, we are interested in the steady state. Firstly, it is trivial to see that $RL$ does not depend on $x^*$. Due to the linearisation that arises from assuming $k_1[L] = \tilde{k}_1$, the stationary distribution for $P(RL)$ is a simple binomial:}
\teoedit{
\begin{equation}
P(RL) = \left( \frac{\tilde{k}_1}{\tilde{k}_1+k_2} \right)^{RL} \left( \frac{k_2}{\tilde{k}_1+k_2} \right)^{R_T-RL} \frac{R_T!}{(R_T-RL)! RL!}, 
\end{equation}
with mean $\langle RL \rangle/R_T = p = \tilde{k}_1 /(\tilde{k}_1+k_2)$. Finding the full solution, $P(x^*, RL) = P(x^*|RL) P(RL)$, is more challenging due to non-linearities that remain in the reactions involving phosphorylation and dephosphorylation of $x$. However, we do note that the behaviour of each readout molecule is independent of the other readouts,  and thus
\begin{equation}
P(x^* | RL) = q(*|RL)^{x^*} (1-q(*|RL))^{x_T- x^*} \frac{x_T!}{x^*! (x_T-x^*)!},
\label{eq:PRL}
\end{equation} 
in which $q(*|RL)$ is the probability that a single readout is phosphorylated in a system with $R_T$ receptors, given that $RL$ of them are ligand-bound. Note that this holds when $RL$ flucuates in time. Even the calculation of $q(*|RL)$ is non-trivial, however, and we do not attempt it here.
}
\teoedit{
\subsection{Mean-field approximation}
To derive Eqs. 11 and 12 of the main text, we make the mean-field assumption that fluctuations in $x^*$ are uncorrelated with fluctuations in $RL$. Mathematically, we assume $P(x^*|RL) = P(x^*|\langle RL \rangle) =P(x^*)$. We combine this assumption with the fact that, as $x^*$ is a 1d variable with reflecting boundary conditions, the total flux associated the increment $x^* \rightarrow x^*+1$ must cancel with the total flux associated with $x^* +1 \rightarrow x^*$. At the mean-field level of approximation, this condition yields
\begin{equation}
\left( k_4 \frac{\langle RL \rangle}{V} + k_6  \frac{R_T - \langle RL \rangle }{V} \right)  (x^*+1) P(x^*+1) = \left(k_3 \frac{\langle RL \rangle}{V} + k_5  \frac{R_T - \langle RL \rangle}{V} \right)  (x_T-x^*)   P(x^*).
\end{equation}
By summing over $x^*$, we obtain the following expression for $\langle x^* \rangle$:
\begin{equation}
\left( k_4 \frac{\langle RL \rangle}{V} + k_6  \frac{R_T - \langle RL \rangle }{V} \right) \langle x^* \rangle = \left(k_3 \frac{\langle RL \rangle}{V} + k_5  \frac{R_T - \langle RL \rangle}{V} \right)  (x_T-\langle x^* \rangle),
\end{equation}
and therefore
\begin{equation}
f = \frac{\langle [x^*] \rangle}{[x_T]} = \frac{k_3 p + k_5 (1-p)}{(k_3+k_4)p + (k_5+k_6)(1-p)},
\end{equation}
as quoted in Eq. 12 of the main text. Further,  the average flux of readouts around the cycle is given by 
\begin{equation}
\dot{n}_{\rm flux} = \frac{k_3}{V} \langle RL x \rangle -  \frac{k_4}{V} \langle RL x^*\rangle = k_3 p R_T [x_T] - \frac{k_3+k_4}{V} \langle RL x^*\rangle.
\end{equation}
In our approximation, $\langle RL x^*\rangle= \langle RL \rangle  \langle x^*|RL\rangle$  and thus
\begin{equation}
\dot{n}_{\rm flux} = (k_3 (1-f) - k_4 f) p [x_T] R_{T},
\end{equation}
as stated in Eq. 11 of the main text. For completeness, we also give the differential equations that govern $RL$ and $x^*$ in the limit of deterministic chemical kinetics:
\begin{equation}
\begin{array}{c}
\frac{{\rm d} RL}{{\rm d} t} = -k_2 RL +\tilde{k}_1 (R_T-RL), \\
\\
\frac{{\rm d} x^*}{{\rm d} t} = -(k_4 [RL] + k_6([R_T]-[RL])) x^*  + (k_3 [RL] + k_5([R_T]-[RL])) (x_T-x^*).
\end{array}
\end{equation}
The steady state of these deterministic equations is given by the stationary stochastic averages in the mean-field approximation: $RL = \langle RL \rangle $ and $x^* = \langle x^*|RL \rangle$.
}

\teoedit{The mean-field limit is valid when the number of receptors is large, so that fluctuations about the average are negligible (note that the number of readouts does not have to be large). The mean-field limit is also valid when the kinetics of receptors is fast compared to that of the readouts --  in this case, $P(x^*|RL) = P(x^*|\langle RL \rangle) =P(x^*)$ since the receptors fluctuate too rapidly to be tracked by the readouts. Indeed, the master equation can be readily solved for a single receptor and a single readout, with the steady-state solution
\begin{equation}
\begin{array}{c}
P(x^*=0, RL=0) = \frac{k_2 (\tilde{k}_1 k_4^\prime + k_2 k_6^\prime+ k_3^\prime k_6^\prime +k_4^\prime k_6^\prime)}{C},\\
\\
P(x^*=1, RL=0) = \frac{k_2 (\tilde{k}_1 k_3^\prime + k_2 k_5^\prime+ k_3^\prime k_5^\prime +k_4^\prime k_5^\prime)}{C},\\
\\
P(x^*=0, RL=1) = \frac{k_1 [L] (\tilde{k}_1 k_4^\prime + k_2 k_6^\prime+ k_4^\prime k_6^\prime +k_4^\prime k_5^\prime)}{C},\\
\\
P(x^*=1, RL=1) = \frac{\tilde{k}_1 (\tilde{k}_1 k_3^\prime + k_2 k_5^\prime+ k_3^\prime k_5^\prime +k_3^\prime k_6^\prime)}{C},
\end{array}
\end{equation} 
in which $k_i^\prime = k_i / V$, and $C$ is a normalizing constant. In the limit $\tilde{k}_1, k_2 \gg k_3^\prime, k_4^\prime, k_5^\prime, k_6^\prime$, the net flux can be directly evaluated as
\begin{equation}
k^\prime_3 P(x^*=0, RL=1) - k_4 P(x^*=1, RL=1) = \frac{(k_3^\prime k_6^\prime - k_4^\prime k_5^\prime) p(1-p)}{p(k_3^\prime + k_4^\prime) + (1-p) (k_5^\prime + k_6^\prime)} = \frac{(k_3 k_6 - k_4 k_5) p(1-p) [x_T] R_T}{p(k_3 + k_4) + (1-p) (k_5 + k_6)}, 
\end{equation}
consistent with Eqs. 11 and 12 of the main text.
}

\teoedit{
Finally, the mean field limit also holds if the receptors do not fluctuate at all -- {\it i.e.}, if $RL$ is fixed over time at a given value. Such a system (which would be a poor concentration sensor) would be equivalent to a system with a fixed number of constitutively active kinases and a fixed number of constitutively active phosphatases. In this case, the fixed value of $RL/R_T$  determines $p$, rather than the average of the ligand binding and unbinding dyamics. However, the equations for $x^*$ and $\dot{n}_{\rm flux}$ hold exactly given this value of $p$. Note that this situation is not equivalent to slow but finite receptor dynamics, in which $RL$ still fluctuates for a given system.
}

\teoedit{
\subsection{Mapping to the copy process does not require the mean-field approximation}
The mapping between an abstract system that performs copies and the biochemical network in Eqs. 3-5 of the main text was performed at the level of transition rates in the Markov model. It is therefore not dependent on a mean-field approximation. We also note that the expression for the average rate at which copies are made,
\begin{equation}
\dot{n}_{\rm copy} = \langle k^{RL}_{\rm copy} x_T +  k^{R}_{\rm copy} x_T  \rangle = \langle\frac{ k_3+k_4}{V} RL x_T +  \frac{ k_5+k_6}{V} R  x_T  \rangle =(( k_3 +k_4) p + (k_5+k_6)(1-p))R_T [x_T],
\end{equation} 
does not rely on a mean-field approximation.
}

\teoedit{
\subsection{Violation of mean field behavior}
\label{violation}
For low numbers of receptors and slow receptor dynamics, the approximation $P(x^*|RL) = P(x^*|\langle RL \rangle) =P(x^*)$ breaks down. In this case, $\langle RL x^* \rangle > \langle RL \rangle \langle x^* |RL\rangle$, since a greater number of ligand-bound receptors leads to more phosphorylation. The result is that  $\dot{n}_{\rm flux}$  is lower than under the mean-field assumption. Since the expression for the rate of copying remains valid,
\begin{equation}
\frac{w_{\rm chem}}{n_{\rm copy}} < \frac{(k_3 k_6 - k_4 k_5) p(1-p)}{\left((k_3+k_4)p + (k_5 + k_6)(1-p)\right)^2} kT \ln \left( \frac{k_3 k_6}{k_4 k_5} \right).
\end{equation}
Physically, the cost of a copy cycle is reduced because the state of the readout is correlated with the receptor prior to copying. This means that fewer actual phosphorylation/dephosphorylation events occur per copy \teoedit{cycle}, and hence less dissipation is necessary. Our comparison to copying, and the calculation of the information generated, assume the measurement of uncorrelated data bits; a lower cost for copying of correlated data bits  can also be achieved with the ideal device in Supplementary Fig. 1. To do this, the decorrelation step 6 in Supplementary Fig. 1 must not be allowed to run to completion -- the barrier between wells should be raised after a certain finite time. The memory bit should then be exposed directly to the next data bit (step 3 in the quasistatic protocol), and the barrier between wells only lowered when the strength of interaction between memory and data bits reflects the correlation between successive data bits. For example, if the correlation between subsequent data bits tends towards unity, the optimal protocol would retain the memory bit in the state of the previous data bit by reducing the time spent in step 6 towards zero. The memory bit would then be exposed to the data bit, but the barrier between wells is not lowered until the coupling is extremely strong. In the limit of perfect correlation, this procedure would require no work. A full exploration of the correlated regime is beyond the scope of this paper, due to its additional complexity. We note in passing that correlated copies of the receptors are not helpful in sensing \cite{govern2014}, and indeed the readout molecules need to respond slowly in order to perform time integration \cite{govern2014}. The regime in which the readout transitions are slower than those of the receptor state not only makes the mean-field analysis accurate, but is also the biologically relevant regime. 
}

\teoedit{
\subsection{Learning rate in the mean-field limit}
The ``learning rate" \cite{Barato2014} gives the rate at which transitions in one part of a bi-partite reaction network act to increase the mutual information between that part of the network and the other part. The most obvious application to a sensory system such as discussed in our manuscript would be to calculate the learning rate that results from the biochemical network's response to changes in the concentration $[L]$. This would be highly relevant to the quality of the network as a sensor of dynamic concentrations. However, it is not of direct relevance here, as we are concerned with whether the action of the readouts can truly be related to computational copying, and if so, how the efficiency relates to thermodynamic bounds on copying. This learning rate would depend both on the steady-state distribution and dynamics of $[L]$.
}
\teoedit{
The receptors and readouts also form a bi-partite system, and hence a learning rate can be defined here. The parameters of the model specify a learning rate, but to calculate it  would require the evaluation of $P(x^*,RL)$. In the mean-field limit we can show that the learning rate is zero.
 The learning rate is \cite{Barato2014}
\begin{equation}
l = -\sum_{x^*,RL} P(x^*,RL) \sum_{x^{*\prime} \neq x^*} w^{RL}_{x^*, x^{*\prime}} \ln \left( \frac{P(x^*,RL)}{P(x^{* \prime},RL)}\right),
\end{equation}
which, using our mean-field approximation, can be re-written as
\begin{equation}
l = -\sum_{x^*} \sum_{x^{*\prime} \neq x^*} P(x^*) W_{x^*, x^{*\prime}} \ln \left( \frac{P(x^*)}{P(x^{* \prime})}\right),
\label{eq:2}
\end{equation}
in which $W_{x^*, x^{*\prime}} = \sum_{RL} P(RL) w^{RL}_{x^*, x^{*\prime}}$. The variable $w_{x^*, x^{*\prime}}$ is only non-zero for adjacent values of $x^{*}$ and $x^{* \prime}$. Further, in the steady state, the average flow from $x^{*}$ to $x^{* \prime}$ must cancel. Therefore we obtain a balance condition that holds for all $x^*$, $x^{*\prime}$:
\begin{equation}
\sum_{RL} P(x^*,RL)  w^{RL}_{x^*, x^{*\prime}}  - P(x^{* \prime},RL)  w^{RL}_{x^{* \prime}, x^{*}} = 0.
\end{equation}
Applying the mean-field approximation to the above expression gives
\begin{equation}
 P(x^*) W_{x^*, x^{*\prime}} =  P(x^{*\prime}) W_{x^{* \prime}, x^{*}}.
\label{eq:4}
\end{equation}
Substituting Eq.\,\ref{eq:4} into Eq.\,\ref{eq:2} gives
\begin{equation}
l =  \sum_{x^*, x^{*\prime}, x^{* \prime} \neq x^*} P(x^*) W_{x^*, x^{*\prime}} \ln \left( {P(x^*)}\right)  -  P(x^{*\prime}) W_{x^{* \prime}, x^{*}} \left({P(x^{* \prime})}\right)=0.
\end{equation}
The fact that the learning rate between receptors and readouts is zero in the mean field limit is consistent with the argument in the previous subsection. In the mean field limit, receptors are perfectly time-averaging, instead of responding to fluctuations; the learning rate reports the degree to which $x^*$ responds to fluctuations in $RL$.
}

\section{Supplementary Discussion 2: Receptors that only act as kinases in the canonical biochemical network}
In the main text  we considered a canonical biochemical system in which receptors were bifunctional -- $RL$ acted as kinases, and $R$ as phosphatases. An alternative is a system in which $RL$ act as kinases but $R$ are inactive, and constitutively active phosphatases exist within the cell. At the level of second-order rate constants, our reactions are:
\begin{equation}
\begin{array}{c}
 RL + x   \underset{k_4}{ \overset{k_3}{\rightleftharpoons}} RL + x^*,\\
 P + x^*  \underset{k_5}{ \overset{k_6}{\rightleftharpoons}} P + x .
\end{array}
\label{reaction rates2}
\end{equation}

In this system, it is not the state of the receptor that is copied into the readout, as $R$ is completely passive. Instead, enzyme identies $RL$ or $P$ are copied into the state of the readout as $x^*$ or $x$ respectively, in the same way that $RL$ and $R$ were copied in the bifunctional system. The relative frequency with which each copy happens can be used to estimate ligand concentration. Eqs. 11 and 12 of the main text still hold, but the \teoedit{average} fraction $f$ of phosphorylated readout \teoedit{in the mean-field limit} is 
\begin{equation}
f = \frac{k_3 p R_T +k_5 P_T}{(k_3+k_4)p R_T + (k_5 + k_6)P_T},
\label{f-equation2}
\end{equation} 
in which $P_T$ is the total number of phosphatases. Thus
\begin{equation}
\dot{w}_{\rm chem} = \frac{(k_3 k_6 - k_4 k_5) p [x_T] P_T R_T}{(k_3+k_4)pR_T + (k_5 + k_6)P_T} kT \ln \left( \frac{k_3 k_6}{k_4 k_5} \right).
\label{G dot2}
\end{equation}
Analogously to the system considered in the main text, the rate at which copies are attempted is
 \begin{equation}
\dot n_{\rm copy2} = [x_T] \left( (k_3+k_4) p R_T + (k_5+k_6) P_T \right).
\label{n dot copy2}
\end{equation}
Combining Equations \ref{G dot2} and \ref{n dot copy2}, the dissipation of free energy per copy \teoedit{cycle} \teoedit{in the mean-field limit} is given by
\begin{equation}
\frac{w_{\rm chem}}{n_{\rm copy}} =  \frac{(k_3 k_6 - k_4 k_5) pP_T R_T}{((k_3+k_4)pR_T + (k_5 + k_6)P_T)^2} kT \ln \left( \frac{k_3 k_6}{k_4 k_5} \right).
\label{work per copy2}
\end{equation}
Further,
\begin{equation}
 p^\prime = p\frac{(k_3+ k_4)R_T}{(k_3+ k_4) pR_T + (k_5 + k_6)P_T}
\end{equation} 
is the probability that an attempted copy is of $RL$ rather than $P$, and 
\begin{equation}
s_{RL} = \frac{k_3}{k_3 + k_4},\,\, s_P= \frac{k_6}{k_5 + k_6}
\label{s1 and s22}
\end{equation}
are the accuracies with which $RL$ and $P$ are copied, respectively.

The results are closely analogous to those obtained in the main text for the canonical biochemical network, and in fact are related by the transformation $(1-p) R_T \rightarrow P_T$ (which is perhaps unsurprising). Indeed, $w_{\rm chem}/n_{\rm copy}$ is the same function of $p^\prime$ in both cases
\begin{equation}
\frac{w_{\rm chem}}{n_{\rm copy}} = (s_1+s_2-1) p^\prime (1-p^\prime) ( E_{s_{RL}}+ E_{s_P}).
\end{equation}
Thus our analysis in the main text still holds, except for the fact that the phosphatase concentration is not directly controlled by varying the probability of receptor-ligand binding $p$. A given cell operating without bi-functional receptors, therefore, does not automatically adapt to high $p$ by converting all phosphatases to kinases, and so will not use a vanishing free-energy per copy \teoedit{cycle} at high $p$ (because $p^\prime$ does not tend towards 1, unlike in the bi-functional system). Its behaviour as a function of $p^\prime$, however, is identical. 

\section{Supplementary Discussion 3: A typical computational  device and protocol}
Here we outline a non-autonomous computational protocol for copying/measurement that reaches the thermodynamic bound of efficiency for a process that performs copies with a given accuracy $s = s_{R} = s_{RL} = 1/(1+\exp(-E_{s}/kT))$, and subsequently allows the information to be lost in an uncontrolled fashion. We shall initially consider a process that reaches the bound when copying data which is in state 1 with a probability $p^\prime=0.5$, before subsequently generalising.

The prototypical memory device is a system capable of storing binary information reliably for a long time: a {\it memory bit}. The state of the  memory bit can be altered by coupling it to external systems; it is important that the memory bit persists in the altered state after the coupling is removed. We shall also manipulate the memory bit using a system of known state, a {\it reference bit}, that allows the memory bit to be reset to a standard state.

The computational device and protocol discussed here is based on that of Bennett \cite{Bennett1982}. \teo{Our memory bit} is  a single particle in a double-well energy landscape, as illustrated in Supplementary Fig.1\cite{Bennett1982,Bennett2003,feynman1998feynman}. A particle in the left well is in state 0, a particle in the right well is in state 1. Coupling the memory bit to an external bit in state 0 raises the right-hand well (stronger coupling implies a larger shift), whereas coupling the memory to an external bit in state 1 raises the left-hand well. 
The barrier between the wells of the memory bit can be lowered and raised as desired.

\subsection{Special case for $p^\prime=0.5$}
The protocol is illustrated in Supplementary Fig. 1. The memory bit starts in state 0 (the left well) with a probability $1/(1+\exp(-E_{ r}/kT))$,
  due to having been equilibrated by a reference bit in state 0 at a coupling strength $E_{ r}$ (the accuracy of this initial reset will turn out to be irrelevant for the cost). The data bit is then copied into the
  memory bit (steps 1-5).  Steps 6-8 return the memory to the standard
  state 0 prior to the next measurement. In step 6, the memory and
  data bits decorrelate, before the actual reset in step 7 and the
  raising of the barrier in step 8. 

To calculate the work done in the protocol, we make the following assumptions. We will assume that the barrier is large enough that hopping between the wells is negligible on all relevant timescales unless the barrier is deliberately lowered. We will also assume that the wells are broad and steep-sided so that the potential energy  of a particle in either well is well-defined. Finally, we assume that raising/lowering the barrier does no work because the probability of finding the particle at the top of the barrier is always negligible (we are considering an ideal two-state system).   For conceptual convenience, we begin our description in state 0, coupled to the reference bit that has been used to reset the memory bit after the last measurement. This choice aligns the start of our cycle with the start of the copy sub-process (the first five stages of the cycle). A reasonable alternative would be to start our description at step 6, 
so that our measurement procedure begins with setting the memory bit to 0 (steps 6-8) and subsequently involves a copy (steps 1-5). This arbitrary choice is of no fundamental importance.  \teoedit{Proceeding step-by-step:}
\begin{enumerate}
\item The barrier is lowered; no work is done.
\item The right well is slowly (quasistatically) lowered from $E_{ r}$ to $0$ as the coupling of the memory bit to a reference bit of known state 0 is reduced. Here, slowness is measured relative to the relaxation time of the system; it is assumed that at any instant, the memory is in a state that is representative of equilibrium given the instantaneous energy landscape. The average work done in this protocol is \cite{Sekimoto1998,Seifert2012}
\begin{equation}
w_2 = \int_{E_{ r}}^0 \frac{e^{-E'/kT}}{1+e^{-E'/kT}} dE' =  -kT \ln \left( \frac{2}{1+e^{-E_{ r}/kT}} \right),
\end{equation}
which tends to $-kT \ln 2 $ in the limit $E_{\rm r}/kT \rightarrow  \infty$. This result shows why the presence of the reference bit at the start of the copy is necessary. \teoedit{Had we instead performed steps 1 and 2 in reverse, lowering the right well to 0 and then lowering the barrier (equivalent to initiating the measurement without a reference bit present),  we would have allowed the memory bit to relax freely to the 50/50 state. The work extracted would have been zero, increasing the total work done during a cycle (by $kT \ln 2$ as $E_r/kT \rightarrow \infty$)}.

\item The memory is exposed to the data bit; as the coupling of the memory bit to the data bit increases, either the left or right well rises from 0 to $E_s$ depending on the state of the data bit. The average work done by slowly coupling the memory to the data bit is
\begin{equation}
w_3 = \int_{0}^{E_s} \frac{e^{-E'/kT}}{1+e^{-E'/kT}} dE' =  kT \ln \left( \frac{2}{1+e^{-E_s/kT}} \right),
\end{equation}
which tends to $kT \ln 2 $ in the limit $E_s/kT \rightarrow  \infty$.
\item The barrier is raised; no work is done
\item The system is decoupled from the data bit, lowering the well that was previously raised. If $E_s$ is finite, there is a finite probability of the particle being in this ``wrong" well; the work done on the system is
\begin{equation}
w_5 = -\frac{e^{-E_s/kT}}{1+e^{-E_s/kT}} E_s,
\end{equation}
which tends to 0 in the limit $E_s/kT \rightarrow  \infty$. This subprocess completes the copying stage of the measurement protocol; the remaining stages involve setting the memory bit to 0 ahead of the next measurement. The probability of performing a correct copy is  $1/(1+\exp(-E_{s}/kT))=s$, as required.

\item  The barrier is lowered. The memory bit  becomes decorrelated from the data bit. The total entropy of memory and data bits therefore increases, but no work is done. In the limit $E_0 \rightarrow \infty$, the memory and data bits are perfectly correlated prior to this step. Therefore the possible configurations of the memory and data bits change from (0,0) or (1,1) to (0,0), (0,1), (1,0) or (1,1), an entropy increase of $k \ln 2$. 
\item The memory is reset by slowly coupling to a reference bit of state 0. The average work done is
\begin{equation}
w_7 = \int_{0}^{E_r} \frac{e^{-E'/kT}}{1+e^{-E'/kT}} dE' =  kT \ln \left( \frac{2}{1+e^{-E_r/kT}} \right),
\end{equation}
which tends to $kT \ln 2 $ in the limit $E_r/kT \rightarrow  \infty$.
\item The barrier is raised; no work is done. The memory bit has been reset to state 0 with probability $1/(1+\exp(-E_{ r}/kT))$.
\end{enumerate} 
\teoedit{The total work done during this cycle is 
\begin{equation}
w_{\rm T} = kT \ln \left( \frac{2}{1+e^{-E_s/kT}} \right) -\frac{e^{-E_s/kT}}{1+e^{-E_s/kT}} E_s.
\label{eq:total work}
\end{equation}
}
As the state of the system is unchanged at the end of the cycle, this work is dissipated irreversibly. If $p^\prime=1/2$, and the accuracy of copying is $s_R = s_{RL} = s$, the information generated by the copy process is
\begin{equation}
I =  s \ln (2s) +  (1-s) \ln(2(1-s)) = \ln(2s)  + (1-s) \ln \left(\frac{1-s}{s}\right) = w_{\rm T}/kT.
\end{equation}
The final equality follows from the definition of $E_s$ in terms of $s$, $E_s = kT \ln(s/(1-s))$ or equivalently $s =1/(1+\exp(-E_{s}/kT))$. Thus the protocol reaches the thermodynamic bound on dissipation for copying data with $p^\prime=1/2$ at an accuracy of $s$.

In the main text, we contrasted the cost of the biochemical network and the optimal protocol for $p^\prime=0.5$, $s_r= s_{RL}=s$ by considering the (free-) energy drops associated with the transitions of the memory. In fact, this is the heat associated with the transitions (or the total drop in chemical free energy of the entire system). One can derive the same cost per copy (Eq.\,\ref{eq:total work}) for an optimal protocol by considering this heat transfer instead of work. For $p^\prime=0.5$, $s_r= s_{RL}=s$, the net heat transfer is given entirely by the heat transferred in step 4; steps 2 and 7 cancel, and the heat is zero for other steps. This is why the intuitive discussion in the main text, which only considers the actual transitions of the memory by which it is set to match the data, is valid.

\subsection{Generalising to $p^\prime \neq 0.5$}

It is clear from the fact that Eq.\,\ref{eq:total work} does not depend on the distribution of input data that the above protocol is not optimal for all $p^\prime$. In this protocol, we made the implicit decision to perform decorrelation (step 6) in a state in which there is no bias between the two wells. We could, instead, consider performing decorrelation by first coupling the memory to another reference bit in state 1 with coupling strength $E_{\rm off}$, then lowering the barrier to allow decorrelation, and finally quasistatically reducing the offset to 0 prior to proceeding with the reset.

There is no change to stages 1-5 or 7-8, in which the total work performed is
\begin{equation}
w_{1-5} + w_{7-8} = kT \ln \left(\frac{2}{1+ \exp(-E_s/kT)}\right). 
\end{equation}
It is now necessary to raise left well (state 0) by an energy $E_{\rm off}$, perform the decorrelation, lower the left well back to $0$ and finally raise the right well to $E_s$.
The work done in raising the left well is
\begin{equation}
w_{6{\rm a}} = (1-f) E_{\rm off}, 
\end{equation}
since a particle will be in the left well after the measurement with a probability $1-f$ by definition ($f=p^\prime s + (1-p^\prime)(1-s)$ is the probability of a measurement outcome being 1). No work is done during the decorrelation stage, and in lowering the left well quasistatically to 0 the work done is
\begin{equation}
w_{6{\rm c}} = \int_{E_{\rm off}}^0 {\rm d}E \frac{\exp(-E/kT)}{1+ \exp(-E/kT)}
=-kT \ln \left(\frac{2}{1+ \exp(-E_{\rm off}/kT)}\right). 
\end{equation} 
Summing over all contributions gives a total work during the cycle of
\begin{equation}
w_{\rm T} = kT \ln \left( \frac{1+ {\rm e}^{-E_{\rm off}/kT}}{1+e^{-E_s/kT}} \right)
 - \frac{{\rm e}^{-E_s/kT}}{1+{\rm e}^{-E_s/kT}} E_s +(1-f) E_{\rm off}.
\end{equation}
At fixed $p^\prime$ and $s$ (and hence $f$), this expression is minimised by
\begin{equation}
1-f - \frac{ {\rm e}^{-E^\dagger_{\rm off}/kT}}{1+e^{-E_{\rm off}/kT}} = 0.
\end{equation}
Hence the optimal decorrelation energy is 
\begin{equation}
E^\dagger_{\rm off} = kT \ln(f/(1-f)),
\end{equation}
for which the work is
\begin{equation}
w^\dagger_{\rm T} = kT \left( s \ln s + (1-s)\ln(1-s) \right) 
- kT \left( f \ln f + (1-f) \ln(1-f)  \right).
\label{work}
\end{equation}
In other words, decorrelation should be performed at a bias that reflects the measurement outcome. We can show that this optimal $E^\dagger_{\rm off}$ saturates the thermodynamic bound; in this case,  the mutual information generated by the copy is
\begin{equation}
I = p^\prime s \ln \left(\frac{s}{f} \right) + p^\prime(1-s) \ln \left( \frac {1-s}{1-f} \right) 
+ (1-p^\prime) s \ln \left(\frac{s}{1-f} \right) + (1-p^\prime)((1-s) \ln \left( \frac {1-s}{f} \right).
\end{equation}
Using $f = p^\prime s + (1-p^\prime) (1-s)$, 
 \begin{equation}
I =  \left( s \ln s + (1-s)\ln(1-s) \right)
- \left( f \ln f + (1-f) \ln(1-f)  \right).
\label{info}
\end{equation}
Comparing Eq. \ref{info} with Eq. \ref{work} shows that $w^\dagger_{\rm T} = kT I$, and therefore that the protocol with appropriately chosen $E^\dagger_{\rm off}$ reaches the thermodynamic bound for copy cycles if the resultant correlations are not used to extract work. 

We note that raising the left well by a negative value of $E_{\rm off}$ to perform decorrelation, in this picture, can only mean raising the right well by $-E_{\rm off}$ by coupling the system to a reference bit in state 0. An analogous derivation to that presented above holds for this operation, so the result holds for both positive and negative $E_{\rm off}$. In this analysis, we haven't considered unequal $s_R$ and $s_{RL}$; it is not obvious how to adjust this protocol to achieve unequal accuracy.

\subsection{Cause of dissipation in the optimal protocol}
\teoedit{As mentioned in the main text, it is actually the decorrelation step that is thermodynamically irreversible, rather than resetting. This is particular clear since the reset energy $E_r$ does not appear in the thermodynamic cost. The cause of irreversibility is the failure to extract work from the combined memory and data bit system as it relaxes from a state more likely to be (memory, data) =  (0,0) or (1,1) than (0,1) or (1,0) to one equally likely to be in (0,0), (0,1), (1,0) or (1,1).  This is directly analogous to the irreversibility of the free expansion of an ideal gas. A procedure that did extract the maximum work from this decorrelation would be fully reversible, and the overall entropy change of the universe during the cycle would be zero, corresponding to the best possible performance of Maxwell's demon \cite{Bennett1982}. Reversing the copying procedure using the same data bit, or ``uncopying" \cite{Ladyman2007}, uses the initial correlation between bits to provide the necessary work to reset the memory bit and is therefore an example of extracting work during decorrelation.}

\teoedit{Not only is resetting reversible, it is unnecessary in an optimal cycle. Steps 1 and 2 are the reverse of steps 7 and 8: the initial part of the measurement protocol reverses the reset to the standard state of 0. Thus these steps are unneccesary, and a cycle of steps 3-6 constitutes an equally valid measurement or copy protocol. As the work required for 7 and 8 exactly cancels the work returned by steps 1 and 2, the total work done in this abridged protocol is still given by Eq. \ref{eq:total work}.
}

\section{Supplementary Discussion 4: sampling with unequal accuracies}

Supplementary Figure 2 shows that $\eta \leq \frac{1}{2}$ for all $p^\prime$ if $s_R = s_{RL} = s$.
In Supplementary Figure 3\,(A-C), we plot efficiency $\eta$ as a function of independently varying $s_{R}$ and $s_{RL}$ for $p^\prime = 0.5,\,0.8,\,0.98$. As $p^\prime$ becomes more extreme, a region of $\eta > \frac{1}{2}$ appears and grows, with the maximal value of $\eta$ found at $s_R \approx 0.5$ and $s_{RL} \approx p^\prime$. Indeed, if we plot $\eta$ against $p^\prime$ with $s_R =0.5$ and $s_{RL} = p^\prime$, as in Supplementary Figure 2\,(D), we see $\eta$ approaching unity as $p^\prime, s_{RL} \rightarrow 1$. For $\eta$ to get close to unity, however, requires extreme values of the parameters. We also note that $\eta \rightarrow 1$ does not occur when the dissipation in a phosphorylation-dephosphorylation cycle is low, and is independent of the absolute reaction rates. 

In the main text, we gave an intuitive explanation of why an efficiency of 0.5 is limiting in the case $s_R=s_{RL}$, $p^\prime =0.5$. This was based on considering the heat deposited into the environment (or the total drop in chemical free energy) by the actual transition of the memory bit. A similar analysis applies here, but when $p^\prime \approx s_{RL} \rightarrow 1$ and $s_R \approx 0.5$, the largest contribution to the overall cost comes from copy cycles involving $RL$ rather than $R$. For the biochemical network, measuring $RL$ involves $s_{RL}-f$ transitions per measurement with a free-energy jump of $E_{s_{RL}}$. For the optimal protocol, the key process is now raising the left well quasistatically from $E_{\rm off}^\dagger$ to $E_{s_{RL}}$, resulting in a transfer of $s_{RL}-f$ particles at an average energy jump of $E_{\rm off}^\dagger < E < E_{s_{RL}}$. When $p^\prime \approx s_{RL} \rightarrow 1$ and $s_R \approx 0.5$, and unlike the  $s_R=s_{RL}$, $p^\prime =0.5$ case, $E_{\rm off}$ and $E_{s_{RL}}$ are both large and similar in magnitude. This fact means that the deposited heat is close to 
  $E_{s_{RL}}$ and hence the efficiency is close to unity.

\section{Supplementary Discussion 5: A clocked biochemical device and protocol}
We  illustrate a discrete biochemical copy process in Supplementary Figure 4. The readout starts in a state $x$ or $x^*$ determined by the previous measurement.
It is then brought into proximity with a receptor (uncorrelated with the previous one) so that
  catalysis can occur.  The readout is allowed to
  reach a steady state, then removed from the proximity of the receptor.

In this protocol the data and memory are allowed to interact for some period of
time and the final result is taken to be the output of a single
copy. The receptor state should remain constant in this period (in practice, constitutively active kinases and
phosphatases could be used). By
  contrast, in the cellular network it is most natural to consider a receptor 
that switches rapidly on timescale of readout
  modification. Nonetheless, as we show next, provided that receptors are
  uncorrelated from one measurement to the next and are presented to readouts 
with the appropriate relative frequency, the cellular and
   clocked biochemical protocols have the same measurement outcomes and efficiency.

If a series of measurements are performed on uncorrelated receptors
using the clocked protocol, the same
phosphorylation fraction $f$ as in the cellular network (Eq. 12 of the main text) will be obtained, provided
that receptors in the ligand-bound state are present
with a probability $p^\prime$. The accuracy of copies is still given
by Eq. 7 of the main text, because $s_R$ and $s_{RL}$ quantify the steady
state probabilities of finding $x^*$ when the receptor is in state
$RL$ and $x$ when the receptor is in state $R$, respectively. The work
per copy \teoedit{cycle} is also given by
Eq. 16 of the main text; to see this, note that the free
energy of the system changes when the molecule $x$ is converted into
$x^*$ or vice versa, either by $RL$ or $R$; \teoqqq{reactions that
  overwrite $x$ with $x$ or $x^*$ with $x^*$ do not cost free energy
  and hence do not contribute to the chemical work.}  The readout $x$
is converted into $x^*$ by $RL$ with probability
$P_{x \rightarrow x^*; RL} = p^\prime s_{RL} (1-f)$
and an associated chemical work of $E_{S_{RL}}$. Similarly, 
$P_{x \rightarrow x^*; R} = (1-p^\prime) (1-s_{R}) (1-f)$, 
$P_{x^* \rightarrow x; RL} = p^\prime (1-s_{RL}) f$, and
$P_{x^* \rightarrow x; R} = p^\prime s_{R} f$,
with associated chemical work $- E_{S_R}$, $-E_{S_{RL}}$ and $E_{S_R}$ respectively.
Using $f = p^\prime s_{RL} + (1-p^\prime)(1-s_R)$ and summing  gives a dissipation per copy \teoedit{cycle} equal to Equation 16 of the main text.

Our clocked biochemical protocol connects to 
 typical computation models that involve manipulating 
energy landscapes, as
illustrated in Supplementary Figure 4 (B).  In effect,
exposing a readout to a receptor in the $RL$ state immediately biases
the landscape towards the $x^*$ ``well'' and simultaneously lowers the
barrier for the transition. Similarly, exposing the readout to $R$ can be interpreted as
suddenly raising the $x^*$ ``well'' and lowering the
barrier. The fundamental difference between  Supplementary Figure 4 (B)
and a optimal computational protocol, such as that in Supplementary Figure 1,  is the
manner in which the landscape switches suddenly, rather than quasistatically.

\section{Supplementary Discussion 6: A thermodynamically optimal copy operation using biomolecules}
\subsection{Detailed calculation for an optimal biochemical bit}
The readout begins coupled to a buffer with with $\Delta G_p, \Delta G_d= -\Delta G_r$ ($\Delta G_r$ assumed to be large and positive, although in fact this initial condition does not influence the copy accuracy or work). There is no receptor in close proximity, but the readout has equilibrated \teoqq{at the end of a previous measurement cycle} by a receptor in the $R$ state, and therefore is in state $x$ with probability $1/(1+ \exp(-\Delta G_r/kT))$.
\begin{enumerate}
\item The readout is brought into close proximity with a receptor of known state $R$; no reactions take place on average. 
\item  $\Delta G_p, \Delta G_d$ are slowly (quasistatically) raised from $-\Delta G_r$ to 0. The chemical work is given by the average number of reactions of a given type, multiplied by the associated chemical work, integrated over the whole process. In the presence of a receptor in state $R$, only  $ R + x^*  {\rightleftharpoons} R + x + {\rm P}$ is possible, and in an infinitesimal step from $ \Delta G_d$ to $\Delta G_d + {\rm d}\Delta G_d$, the average number of net dephosphorylation events is
\begin{equation}
n_{\rm dephos} = \frac{\exp{(-\Delta G_d/kT)}}{1+\exp{(-\Delta G_d/kT)}} - \frac{\exp{(-(\Delta G_d+{\rm d}\Delta G_d) /kT)}}{1+\exp({-(\Delta G_d +{\rm d}\Delta G_d)/kT})} 
= - \frac{\exp{(-\Delta G_d/kT)}}{\left(1+\exp{(-\Delta G_d/kT)}\right)^2} \frac{{\rm d}\Delta G_d}{kT}.
\end{equation}
Thus
\begin{equation}
w_{2,{\rm chem}} =  \int_{-\Delta G_r}^0   \frac{{\rm d}\Delta G_d}{kT}   \frac{(\Delta G_d +  \Delta G_{x/x^*}) \exp{(-\Delta G_d/kT)}}{\left(1+\exp{(-\Delta G_d/kT)}\right)^2},
\end{equation}
since $-(\Delta G_d +  \Delta G_{x/x^*})$ is the chemical work done by the buffer in a single dephosphorylation reaction catalysed by $R$. Computing the integral,
\begin{equation}
w_{2,{\rm chem}} =  -kT \ln \left(\frac{2}{1+ \exp(-\Delta G_r/kT)} \right) +  \frac{\Delta G_r \exp(-\Delta G_r/kT)}{1+ \exp(-\Delta G_r/kT)} 
-\Delta G_{x/x^*} \left( \frac{1}{2} -\frac{1}{1+\exp(-\Delta G_r/kT)} \right).
\label{eq:wc2}
\end{equation}
\item The readout is brought into close proximity with a receptor of unknown state (ligand-bound with probability $p^\prime$). $\Delta G_p, \Delta G_d$ are then slowly lowered to $-\Delta G_s$. In this step, the state of the readout is set to match that of the receptor (with some error). Proceeding analogously to Step 2, and considering the possibility of the receptor being either $R$ or $RL$ with probabilities $1-p^\prime$ and $p^\prime$, respectively,
\begin{equation}
\begin{array}{c}
w_{3,{\rm chem}} =  +kT \ln \left(\frac{2}{1+ \exp(-\Delta G_s/kT)} \right) -  \frac{\Delta G_s \exp(-\Delta G_s/kT)}{1+ \exp(-\Delta G_s/kT)} \\
\\
+(1-2p^\prime)\Delta G_{x/x^*} \left( \frac{1}{2} -\frac{1}{1+\exp(-\Delta G_s/kT)} \right).
\end{array}
\label{eq:wc3}
\end{equation}
\item The readout is removed from close proximity with the receptor. No reactions occur on average.
\item $\Delta G_p, \Delta G_d$ are set to $\Delta G_{\rm off}$. No reactions occur on average. This stage constitutes the end of the copy; if the unknown receptor is in state $RL$, then the readout is in $x^*$ with probability $s = 1/(1+\exp(-\Delta G_s/kT))$. Similarly, if the unknown receptor is in state $R$, the readout is in $x$ with probability $s = 1/(1+\exp(-\Delta G_s/kT))$. Thus the protocol performs a copy of accuracy $s$.
\item We now decorrelate the memory and data bits. The readout is brought into close proximity with a known receptor of state $R$. The readout relaxes to a state reflective of  $\Delta G_{\rm off}$ via the reaction $ R + x^*  {\rightleftharpoons} R + x + {\rm P}$. The chemical work done is
\begin{equation}
\begin{array}{c}
w_{6,{\rm chem}} = - \left(\Delta G_{\rm off} + \Delta G_{x/x^*} \right) \times 
p^\prime \left( \frac{ 1}{1+ \exp(-\Delta G_s/kT)} -  \frac{1}{1+ \exp(-\Delta G_{\rm off}/kT)} \right) \\
\\
+ \left(\Delta G_{\rm off} + \Delta G_{x/x^*} \right) \times 
(1-p^\prime) \left( \frac{1}{1+ \exp(-\Delta G_s/kT)} +  \frac{1}{1+ \exp(-\Delta G_{\rm off}/kT)}-1 \right). 
\end{array}
\label{eq:wc6}
\end{equation}
\item The readout molecule is reset by quasistatically lowering $\Delta G_p, \Delta G_d$ to $-\Delta G_r$ from $\Delta G_{\rm off}$, returning it to a state dominated by $x$.
\begin{equation}
\begin{array}{c}
w_{7,{\rm chem}} =  +kT \ln \left(\frac{1+ \exp(\Delta G_{\rm off}/kT)}{1+ \exp(-\Delta G_r/kT)} \right)
 -  \frac{\Delta G_r \exp(-\Delta G_r/kT)}{1+ \exp(-\Delta G_r/kT)} -\frac{\Delta G_{\rm off} \exp(\Delta G_{\rm off}/kT)}{1+ \exp(\Delta G_{\rm off}/kT)}\\
\\
+ \Delta G_{x/x^*} \left(1- \frac{1}{1+ \exp(-\Delta G_{\rm off}/kT)} -\frac{1}{1+\exp(-\Delta G_r/kT)} \right).
\end{array}
\label{eq:wc7}
\end{equation} 
\item The readout molecule is separated from the known receptor. No reactions occur on average. The system has now been returned to the initial state.
\end{enumerate}
Summing Eqs. \ref{eq:wc2}, \ref{eq:wc3}, \ref{eq:wc6} and \ref{eq:wc7} gives a total chemical work for the cycle of 
\begin{equation}
w_{\rm chem} =  kT \ln \left(\frac{1+\exp(\Delta G_{\rm off}/kT)}{1+ \exp(-\Delta G_s/kT)} \right) -  \frac{\Delta G_s \exp(-\Delta G_s/kT)}{1+ \exp(-\Delta G_s/kT)} 
- \Delta G_{\rm off} \frac{p^\prime + (1-p^\prime) \exp(-\Delta G_s/kT)}{1+ \exp(-\Delta G_s/kT)}.
\label{w1}
\end{equation}
Using the fact that the probability of the copy resulting in $x^*$ (at the end of step 5) is  $f = p^\prime s + (1-p^\prime)(1-s)$, the fact that the accuracy $s$ is given by $s=1/(1+\exp(-\Delta G_s/kT))$, and the identity $\exp(z) \equiv (1+ \exp(z))/(1+\exp(-z))$, 
\begin{equation}
w_{\rm chem} =  kT \ln \left(\frac{1+\exp(-\Delta G_{\rm off}/kT)}{1+ \exp(-\Delta G_s/kT)} \right)  - (1-s) \Delta G_s
+(1-f) \Delta G_{\rm off}.
\end{equation}
Similarly to Supplementary Discussion 3, we can minimize the work with respect to $\Delta G_{\rm off}$ at fixed $p^\prime$, $s$ and $f$, obtaining $\Delta G_{\rm off}^\dagger = kT \ln (f/(1-f))$. Substituting for $\Delta G_{\rm off}^\dagger$ and $\Delta G_s$ using $f$ and $s$, we see that this particular protocol involves work
\begin{equation}
w_{\rm chem} =  kT \left( s \ln s + (1-s) \ln(1-s)
- f \ln f -(1-f) \ln(1-f) \right),
\end{equation}
which is identical to the information gained by the measurement of the data (Supplementary Eq.\,\ref{info}). Thus the biochemical protocol with decorrelation free energy $\Delta G_{\rm off}^\dagger = kT \ln (f/(1-f))$ is a thermodynamically optimal protocol.

\subsection{Subtleties in the comparison between biochemical systems and energy landscapes}
\subsubsection{The role of receptors}
In the energy-landscape description presented in Supplementary Figure 1, it was assumed that we are separately able to lower the barrier between wells and raise one well in energy with respect to the other.  In the quasistatic biochemical protocol presented in the main text, the receptor acts as a catalyst (lowering the barrier between activation states of the readout), but the state of the receptor also determines which state is favourable. Thus it is not possible to formally separate biasing and lowering/raising of the barrier in the same way as in the energy landscape picture (the strength of the bias when a receptor is present, but not its direction, is determined by the buffer to which the memory is connected). Indeed, it is not really meaningful to draw energy landscapes in a two-state description of the biochemical system in which the fuel is coarse-grained away, unless there is a single receptor present. As we demonstrate, however, this does not prevent the biochemical system performing efficient measurement, and the purpose of each of the 8 steps in the biochemical protocol is closely analogous to its counterpart in Supplementary Figure 1.

\subsubsection{Chemical work}
We claim that there is a close analogy between the optimal biochemical protocol in the main text and the energy-landscapes description in Supplementary Figure 1. Further, we claim that the purpose of the individual steps are the same. However, the work calculated for the individual steps  are quite different; for example, the ``chemical work" during the decorrelation step (step 6) in the biochemical protocol is non-zero, whereas the work done is zero in the energy landscapes picture. 

The work for the protocol in Supplementary Figure 1  is calculated in  Supplementary Discussion 3 as  is typical in stochastic thermodynamics \cite{Seifert2011}. Work is done when the energy landscape is manipulated, and heat is exchanged when the particle moves within the landscape. Although this approach is well-established for such systems, it is less clear how to proceed when the external manipulation involves the concentrations of reactants, as in our biochemical approach. Furthermore, as discussed above, the use of (free-) energy landscapes to describe the biochemical system at a coarse-grained level is not straight-forward. 

Instead, we choose to calculate the change in chemical free energy associated with the buffers (the negative of the ``chemical work" they perform). This quantity is well defined, and in a measurement cycle (in which the memory returns to its initial state) corresponds to the overall change in free energy of the entire system. A decrease in free energy of the system corresponds to a reduction of its ability to do work, and is therefore ``dissipation" unless this drop is harnessed to do work.

\section{Supplementary Discussion 7: An optimal protocol involving a receptor that functions only as a kinase}

It is possible to construct a measuring device using only the reaction 
\begin{equation}
RL + x +{\rm ATP} {\rightleftharpoons} RL + x^* + {\rm ADP}
\label{eq:just kinase}
\end{equation}
 if the $R$ state of the receptor does not catalyse a reaction. The procedure is illustrated in Supplementary Fig. 5, and outlined in detail below. As in Supplementary Discussion 6, we calculate the chemical work performed by the ATP/ADP reservoirs during a cycle, in which the unknown receptor is in state $RL$ with probability $p^\prime$. We start with the readout isolated from any receptors, and in a buffer with a high ADP concentration and a low ATP concentration. In this case, the reaction in Eq. \ref{eq:just kinase} is driven to the left; $\Delta G_p = \Delta G_s$ where $\Delta G_s$ is large and positive. The readout has previously been equilibrated by a receptor in the $RL$ state at this value of  $\Delta G_p$; it is therefore predominantly in the $x$ state. Note that since 
the measurement of $R$ will be passive, the initial state determines the accuracy of measurement and hence $\Delta G_s$ appears here, rather than an arbitrary reset accuracy $\Delta G_r$.

\begin{enumerate}
\item The readout is brought into close proximity with the unknown receptor. No reactions occur on average 
\item $\Delta G_p$ is slowly (quasistatically) lowered from  $\Delta G_s$ to $-\Delta G_s$. If the unknown receptor is in state $RL$, the readout tends to be converted to $x^*$ -- no reactions take place if the receptor is in state $R$.  The average chemical work done is 
\begin{equation}
\begin{array}{c}
w_{2,{\rm chem}} =  p^\prime \int_{\Delta G_s}^{-\Delta G_s}   \frac{{\rm d}\Delta G_p}{kT}  \frac{( \Delta G_p - \Delta G_{x/x^*})   \exp{(-\Delta G_p/kT)}}{\left(1+\exp{(-\Delta G_p/kT)}\right)^2},\\
\\
= p^\prime \Delta G_{x/x^*} \left( \frac{1}{1 +\exp{(-\Delta G_s/kT)} }-  \frac{1}{1 +\exp{(\Delta G_s/kT)} }\right).  
\end{array}
\end{equation}
The relation $\exp(x) \equiv (1+ \exp(x))/(1+\exp(-x))$ is useful in deriving this result.
\item The readout is separated from the receptor; no reactions take place on average.
\item The buffers are slowly changed so that $\Delta G_p =-\Delta G_{\rm off}$. No reactions take place. This stage constitutes the end of the copy; if the unknown receptor is in state $RL$, then the readout is in $x^*$ with probability $s = 1/(1+\exp(-\Delta G_s/kT))$. Similarly, if the unknown receptor is in state $R$, the readout is in $x$ with probability $s = 1/(1+\exp(-\Delta G_s/kT))$. Thus the copy has been performed with accuracy $s  = 1/(1+\exp(-\Delta G_s/kT))$.
\item The following steps involve decorrelation and resetting. The readout is brought into contact with the receptor of known state $RL$. The readout relaxes to a state representative of the free-energy difference $\Delta G_p=-\Delta G_{\rm off}$. The chemical work done is
\begin{equation}
\begin{array}{c}
 w_{5,{\rm chem}}=p^\prime  \left( \Delta G_{x,x^*} + \Delta G_{\rm off}  \right)  \left(\frac{1}{1 +\exp{(-\Delta G_{\rm off}/kT)} }-  \frac{1}{1 +\exp{(-\Delta G_s/kT)} } \right)\\
  \\
  + (1- p^\prime)\left( \Delta G_{x,x^*} + \Delta G_{\rm off}  \right)
  \left(\frac{1}{1 +\exp{(-\Delta G_{\rm off}/kT)} }-  \frac{1}{1 +\exp{(\Delta G_s/kT).} } \right).\\
 \end{array}
\end{equation}
\item The buffer is slowly changed from $\Delta G_p = -\Delta G_{\rm off}$ to $\Delta G_p = \Delta G_s$. The readout is restored to a state dominated by $x$. The chemical work done is
\begin{equation}
\begin{array}{c}
w_{6,{\rm chem}} =  \int_{-\Delta G_{\rm off}}^{\Delta G_s}   \frac{{\rm d}\Delta G_p}{kT}  \frac{( \Delta G_p - \Delta G_{x/x^*})   \exp{(-\Delta G_p/kT)}}{\left(1+\exp{(-\Delta G_p/kT)}\right)^2}
=kT \ln \left(\frac{1+ \exp(-\Delta G_{\rm off}/kT)}{1+ \exp(\Delta G_s/kT)} \right) \\
\\
 +  \frac{\Delta G_s \exp(\Delta G_s/kT)}{1+ \exp(\Delta G_s/kT)} +\frac{\Delta G_{\rm off} \exp(-\Delta G_{\rm off}/kT)}{1+ \exp(-\Delta G_{\rm off}/kT)}
+ \Delta G_{x/x^*} \left( \frac{1}{1 +\exp{(\Delta G_s/kT)} }-  \frac{1}{1 +\exp{(\Delta G_{\rm off}/kT)} }\right).  
\end{array}
\end{equation}
\item The readout is removed from the receptor. No reactions take place. The system is restored to the initial state.
\end{enumerate}
Summing over all terms gives
\begin{equation}
w_{\rm chem}  =  kT \ln \left(\frac{1+\exp(-\Delta G_{\rm off}/kT)}{1+ \exp(\Delta G_s/kT)} \right) +  \frac{\Delta G_s \exp(\Delta G_s/kT)}{1+ \exp(\Delta G_s/kT)} 
+\Delta G_{\rm off} \frac{(1-p^\prime) + p^\prime \exp(-\Delta G_s/kT)}{1+ \exp(-\Delta G_s/kT)}.
\label{wcyc method2.0}
\end{equation}
Utilising $\exp(x) \equiv (1+ \exp(x))/(1+\exp(-x))$ once more, The Eq. \ref{wcyc method2.0} can be re-written
\begin{equation}
w_{\rm chem} =  kT \ln \left(\frac{1+\exp(\Delta G_{\rm off}/kT)}{1+ \exp(-\Delta G_s/kT)} \right) -  \frac{\Delta G_s \exp(-\Delta G_s/kT)}{1+ \exp(-\Delta G_s/kT)} 
-\Delta G_{\rm off} \frac{p^\prime + (1-p^\prime) \exp(-\Delta G_s/kT)}{1+ \exp(-\Delta G_s/kT)},
\label{wcyc method2.1}
\end{equation}
which is identical to Supplementary Eq.\,\ref{w1}. The above method therefore has the same outcome and thermodynamic cost as those analysed in Supplementary Discussions 3 and 6 It has the advantage of requiring the manipulation of $\Delta G_p$ only, but the disadvantage that the reset step is actually necessary. The reset step is needed because the $R$ state does not actively set the the memory to $x$; we rely on the memory being prepared in the $x$ state prior to copy. 

\end{widetext}
\end{appendix}

\end{document}